\documentclass[showpacs,preprintnumbers,amsmath,aps,nofootinbib,amssymb,superscriptaddress]{revtex4}
\usepackage{graphicx}
\usepackage[english]{babel}
\usepackage{bm}
\usepackage[font=scriptsize]{caption}
\usepackage{ragged2e}
\usepackage[font=small,labelfont=bf,justification=raggedright]{caption}
\usepackage[utf8]{inputenc}
\usepackage[tikz]{bclogo}
\usepackage[framemethod=tikz]{mdframed}

\newsavebox{\measurebox}
\begin{document}
\baselineskip 16pt
\title{Anisotropic chiral cosmology: exact solutions}
\author{Luis Rey D\'iaz-Barr\'on}
\email{lrdiaz@ipn.mx}
\affiliation{Unidad Profesional
Interdisciplinaria de Ingenier\'ia,
Campus Guana\-jua\-to del Instituto Polit\'ecnico Nacional.\\
Av. Mineral de Valenciana \#200, Col. Fraccionamiento Industrial
Puerto Interior, C.P. 36275, Silao de la Victoria, Guana\-jua\.to,
M\'exico.}
\author{Abraham Espinoza-Garc\'ia}
\email{aespinoza@ipn.mx}
\affiliation{Unidad Profesional
Interdisciplinaria de Ingenier\'ia,
Campus Guana\-jua\-to del Instituto Polit\'ecnico Nacional.\\
Av. Mineral de Valenciana \#200, Col. Fraccionamiento Industrial
Puerto Interior, C.P. 36275, Silao de la Victoria, Guana\-jua\.to,
M\'exico.}

\author{S. P\'erez-Pay\'an}
\email{saperezp@ipn.mx}
\affiliation{Unidad Profesional
Interdisciplinaria de Ingenier\'ia,
Campus Guana\-jua\-to del Instituto Polit\'ecnico Nacional.\\
Av. Mineral de Valenciana \#200, Col. Fraccionamiento Industrial
Puerto Interior, C.P. 36275, Silao de la Victoria, Guana\-jua\.to,
M\'exico.}

\author{J. Socorro}
\email{socorro@fisica.ugto.mx}
 \affiliation{Departamento de
F\'{\i}sica, DCeI, Universidad de Guanajuato-Campus Le\'on, C.P.
37150, Le\'on, Guanajuato, M\'exico}

\begin{abstract}
In this work, we investigate the anisotropic Bianchi type I cosmological model in the chiral setup, in a twofold manner.
Firstly, we consider a quintessence plus a k-essence like model, where two scalar fields but only one potential term is considered.
Secondly, we look at a model where in addition to the two scalar fields the two potential terms are taken into account as well as
the standard kinetic energy and the mixed term. Regarding this second model, it is shown that two possible cases can be studied:
a quintom like case and a quintessence like case. In each of the models, we were able to find both classical and quantum analytical solutions.
\end{abstract}
\keywords{Bianchi I; Multi-Field Cosmology; Exact solutions.}

\pacs{98.80.Qc, 98.80.Es, 04.20.Jb}
 \maketitle
\section{Introduction}

It is well established that our universe is homogeneous and
isotropic at large scales and can be modeled by the flat
Friedman-Lema$\hat\i$tre-Robertson-Walker (FLRW) geometry. However,
observations of the cosmic microwave background (CMB) have shown the
existence of anomalies at large angular scales
\cite{Schwarz:2015cma}, this has led to put forward the hypothesis
of primordial anisotropies in the early stages of the universe and
could shed some light on the anomalies found in the CMB. Therefore,
it is reasonable to address these issuess. Attempts to incorporate
these ideas into the cosmological setting have been presented in
Refs.~\cite{Starobinsky1,Starobinsky2,Starobinsky3,Starobinsky4,Starobinsky5,Pereira:2007yy,Pitrou:2008gk,Pereira:2015pga,Gumrukcuoglu:2007bx},
where anisotropic cosmological models have been used (mostly the
Bianchi I model) as a background space-time in an early anisotropic
but homogeneous universe that develops isotropization at the
beginning of inflation, nonetheless, the imprints of such anisotropy
would lead to the thermal maps of the CMB; and once the inflationary
period ends, as a consequence of this isotropization, the universe
acquires a FLRW geometry recovering the standard picture of the
evolution of the universe. Due to the above, anisotropic
cosmological models represent an attractive arena to test the early
stages of the universe, even if no conclusive evidence that a
primordial anisotropy is needed.

To tackle the different phenomena of our universe, scalar field cosmological models have been broadly used. Some of the issues that have been
addressed under this line of thought are the dark matter component of the universe, the late time acceleration and the inflationary epoch, to name
a few \cite{Starobinsky6,linde1982, Linde1983, Barrow1993_1,Barrow1993_2,Peebles1987, Tsujikawa2013, Liddle1998, Sahni1999, Matos2000,Urena-Lopez2016, Peebles1998, deHaro:2016_1, deHaro:2016_2, Elizalde:2004mq}. Although single scalar field cosmological models have been a cornerstone in giving answers to different problems present in our universe, there are still illnesses that haven't been cured. On this regard, in recent years cosmological models considering two or more scalar fields have drawn tremendous attention. The advantages of these models (compared to the single field ones) is the introduction of new degrees of freedom which allows the explanation of several physical phenomena.

Generally speaking, in this multi-scalar field cosmological models, the interaction of the scalar fields occur in the potential, the mixed kinetic terms or both. In this setup, an inflationary picture of the universe can also be achieved \cite{Coley:1999mj, Copeland:1999cs}, even if the interaction between the scalar fields does not take place \cite{andrew2007}. Moreover, multi-scalar
fields models can also be used to explain the primordial inflation
perturbations analysis \cite{Yokoyama:2007dw, Chiba:2008rp} or the
assisted inflation  \cite{Copeland:1999cs, andrew1998a}. Another
appealing reason to work with these models is that when two scalar fields
are considered, the crossing of the cosmological constant boundary
-1 can be described, in the litearature these models are known as
quintom models \cite{Cai2009, Setare2008, Lazkoz2007,Leon2018}
(single scalar field models do not have this malleability, since
they only describe either the phantom or quintessence regime).
Furthermore, this multi-field models can also tackle the hybrid
inflation of the universe, which gives an alternative graceful exit
in comparison to the standard inflationary picture
\cite{chimento,lindle, cope, kim,omar-epjp2017,Wands2008, Bond2006,
Inomata2017}. The most successful models, phenomenologically speaking, are those that have incorporated quintessence scalar fields,
slow-roll inflation, chiral cosmology connected with
$f(R)$ theories and the nonlinear sigma model \cite{OSR, Liddle1998,
barrow, ferreira, copeland1, copeland2,
copeland3,andrew2007,gomez,capone,kolb,Vagnozzi,chervon1995,Chervon2013,Kaiser2014,Fomin2017,Chervon2019,Paliathanasis2019,Paliathanasis2020a,Paliathanasis2020b,Bamba:2012cp,Dimakis:2020tzc,Dimakis:2019qfs,Paliathanasis:2014yfa}.

In connection with the latter, multi-field anisotropic cosmological
models of inflation have been explored. In
Ref.~\cite{Folomeev:2007uw} the author delves into the study of the
Bianchi type I cosmology considering two interacting scalar fields
and a potential of the form $V(\phi,\chi)\sim\phi^4+\chi^4$,
founding numerical solutions as well as the asymptotically isotropic
Friedmann case. Other interesting works are presented in Refs.
\cite{omar-epjp2017} and \cite{socorro-libro}, where the potential
with structure $V(\phi,\sigma)\sim e^{\phi+\sigma}$ has shown to be
a good viable candidate to address the inflationary era in both flat
isotropic and anisotropic space-times. More recently, in Ref.
\cite{sor} the authors present the case of the anisotropic Bianchi
type I cosmology in the multi-field setup with a potential of the
form $V_0e^{-(\lambda_1\phi_1+\cdots+\lambda_n\phi_n)}$, founding
inflationary exact solutions in a quintessence framework. Additional
research regarding multi-filed anisotropic cosmological models can
be found in Refs.
\cite{Chervon2013,Andrianov:2015hba,GALIAKHMETOV:2014zea,Abbyazov:2013qqa,Chervon:2015jji,Cicciarella:2019ihh,Herfray:2015fpa,Leon:2020pfy,Kaiser:2010ps,Kaiser:2013sna,Beesham:2013rya}.

In the present work we present the anisotropic Bianchi type I
cosmological model with two scalar fields in a twofold manner.
Following closely the developments introduced in Refs.
\cite{Socorro:2019vvh} and \cite{Socorro:2020nsm}, first, we put
forward a simple quintessence plus a k-essence model which arises
from considering the interaction of the two scalar fields but only
one potential term. And second, a chiral approach is studied, in
this case, in addition to the previous two scalar fields we also
consider the two potential terms as well as the standard kinetic
energy and the mixed term. For each model, classical and quantum
analytical solutions are found.

This paper is arranged as follows. In section \ref{first-model} we introduce the first model, where the Einstein-Klein-Gordon (EKG) equations
are calculated and the Lagrangian and Hamiltonian approach is implemented in order to find the corresponding solutions as well as the
anisotropic parameters. In section \ref{second-model}, the second model is presented, here, after obtaining the Hamiltonian density
we can distinguish two possible scenarios: a quintom like case and a quintessence case. For both scenarios the corresponding solutions
are found. Section \ref{quantum-versions} is devoted to implement the quantum versions of the previous two models and the corresponding
solutions are obtained. Finally, section \ref{conclusions} is left for the final remarks.
\section{First model: quintessence plus k-essence}\label{first-model}
As we already mentioned, we are going to start by analyzing the quintessence plus k-essence model. For this purpose let us consider
the Lagrangian density for such a model, which reads
\begin{equation}
\rm {\cal L}=\sqrt{-g} \left( R-\frac{1}{2}g^{\mu\nu} \nabla_\mu
\phi_1 \nabla_\nu \phi_1 -\frac{1}{2}g^{\mu\nu} \nabla_\mu \phi_2
\nabla_\nu \phi_2 + V(\phi_1)\right) \,, \label{first-lagra}
\end{equation}
where $\rm R$ is the Ricci scalar, $\rm V(\phi_1)=V_1e^{-\lambda_1\phi_1}$ is the
corresponding scalar field potential (as it will be shown below),  and the reduced Planck mass
$M_{P}^{2}=1/8\pi G=1$. The corresponding variations of (\ref{first-lagra}), with respect to the metric and the scalar
fields give the EKG field equations
\begin{eqnarray}
\rm G_{\alpha \beta} &=&\rm -\frac{1}{2} \left(\nabla_\alpha \phi_1
\nabla_\beta \phi_1 -\frac{1}{2}g_{\alpha \beta} g^{\mu \nu}
\nabla_\mu \phi_1 \nabla_\nu \phi_1 \right) +\frac{1}{2}g_{\alpha \beta} \, V(\phi_1)\nonumber\\
&&\rm -\frac{1}{2} \left(\nabla_\alpha \phi_2 \nabla_\beta \phi_2
-\frac{1}{2}g_{\alpha \beta} g^{\mu \nu}
\nabla_\mu \phi_2 \nabla_\nu \phi_2 \right), \label{munu}\\
\rm \Box \phi_1 -\frac{\partial V}{\partial \phi_1} &=&\rm
g^{\mu\nu} {\phi_1}_{,\mu\nu} - g^{\alpha \beta} \Gamma^\nu_{\alpha
\beta} \nabla_\nu \phi_1 - \frac{\partial V}{\partial \phi_1}=\rm 0
\,,\label{ekg-phi-1}\\
&& \rm g^{\mu\nu} {\phi_2}_{,\mu\nu} - g^{\alpha \beta}
\Gamma^\nu_{\alpha \beta} \nabla_\nu \phi_2=0.\label{ekg-phi2}
\end{eqnarray}
The line element for the anisotropic cosmological Bianchi type I model in the Misner parametrization is
\begin{eqnarray}
\rm ds^2 &=& \rm -N^2 dt^2 +a_1^2 dx^2 + a_2^2 dy^2 + a_3^2 dz^2, \nonumber \\
 &=&\rm -N^2 dt^2 + e^{2\Omega}\left[e^{2\beta_++2\sqrt{3}\beta_-}dx^2 +e^{2\beta_+-2\sqrt{3}\beta_-} dy^2 + e^{-4\beta_+} dz^2\right],
 \label{biachi_I_misner}
 \end{eqnarray}
 where $\rm a_i$ ($\rm i=1,2,3$) are the scale factors in directions $\rm (x,y,z)$, respectively, and N is the lapse function.
 For convenience, and in order to carry out the analytical calculations, we consider the following representation for the line
 element (\ref{biachi_I_misner})
\begin{equation}
\rm ds^2 = -N^2 dt^2 + \eta^2\left[m_1^2 dx^2 +m_2^2 dy^2 + m_3^2
dz^2\right], \label{bianchi}
 \end{equation}
 where the relations between both representations (\ref{biachi_I_misner}) and (\ref{bianchi}) are given by
 \begin{eqnarray}
 \rm \eta&=& \rm e^{\Omega}, \nonumber\\
 \rm m_1 &=& \rm e^{\beta_+ + \sqrt{3} \beta_-}, \qquad \frac{\dot m_1}{m_1}=\dot \beta_+ + \sqrt{3} \dot \beta_-, \nonumber\\
 \rm m_2&=& \rm e^{\beta_+ - \sqrt{3} \beta_-}, \qquad \frac{\dot m_2}{m_2}=\dot \beta_+ - \sqrt{3} \dot \beta_-,\\
 \rm m_3&=& \rm e^{-2\beta_+ }, ~~\quad\qquad \frac{\dot m_3}{m_3}=-2\dot \beta_+,\nonumber
\end{eqnarray}
and $\eta$ is a function that has information regarding the isotropic scenario and the $\rm m_i$ are dimensionless functions
 that have information about the anisotropic behavior of the universe, such that
\begin{equation}
\rm  \prod_{i=1}^{3}m_i=1, \qquad
 \prod_{i=1}^{3}a_i= \eta^3, \qquad
\rm  \sum_{i=1}^3 \frac{\dot m_i}{m_i}= 0, \label{mi}
\end{equation}
act as constraint equations for the model.

\subsection{General Solutions to the Field Equations}
In this subsection we present the solutions of the field equations for
the anisotropic cosmological model, considering the temporal
evolution of the scale factors with barotropic fluid and standard
matter. The solutions obtained already consider the particular
choice of the Misner-like transformation discussed lines above.
Using the metric (\ref{bianchi}) and a co-moving fluid, equations
(\ref{munu}) take the following form
\begin{eqnarray}
&&\rm\frac{\dot m_1}{Nm_1} \frac{\dot m_2}{Nm_2}+\frac{\dot m_2}{Nm_2}\frac{\dot m_3}{Nm_3}+\frac{\dot m_1}{Nm_1}\frac{\dot m_3}{Nm_3}+3 \left(\frac{\dot \eta}{N\eta}\right)^2-8\pi G\rho \label{0,0}\\
&&\rm\mbox{} -\frac{1}{2}\left(  \frac{1}{2} \frac{\dot \phi_1^2}{N^2}+V(\phi_1)\right)-\frac{1}{4} \frac{\dot \phi_2^2}{N^2}=0,\nonumber\\
&&\rm
-\frac{\dot N}{N^2}\left[\frac{\dot m_2}{Nm_2}+\frac{\dot m_3}{Nm_3}+2\frac{\dot \eta}{N\eta}\right]+\frac{\ddot m_2}{N^2m_2}+ \frac{\ddot m_3}{N^2m_3}+\frac{\dot m_2}{Nm_2}\frac{\dot m_3}{Nm_3}+2\frac{\ddot \eta}{N^2\eta}\label{1,1}\\
&&\mbox{}\rm+\left(\frac{\dot \eta}{N\eta}\right)^2 +3\frac{\dot \eta}{N\eta}\left[\frac{\dot m_2}{Nm_2}+\frac{\dot m_3}{Nm_3} \right] \rm +\frac{1}{2}\left(  \frac{1}{2} \frac{\dot\phi_1^2}{N^2}-V(\phi_1)\right)+\frac{1}{4} \frac{\dot \phi_2^2}{N^2}+8\pi G P=0 \nonumber,\\
&&\rm
 -\frac{\dot N}{N^2}\left[\frac{\dot m_1}{Nm_1}+\frac{\dot m_3}{Nm_3}+2\frac{\dot \eta}{N\eta}\right]+\frac{\ddot m_1}{N^2m_1} + \frac{\ddot m_3}{N^2m_3}+\frac{\dot m_1}{Nm_1}\frac{\dot m_3}{Nm_3}+2\frac{\ddot \eta}{N^2\eta}\label{2,2}\\
&&\mbox{}\rm+\left(\frac{\dot \eta}{N\eta}\right)^2 +3\frac{\dot \eta}{N\eta}\left[\frac{\dot m_1}{Nm_1}+\frac{\dot m_3}{Nm_3} \right] \rm   +\frac{1}{2}\left(  \frac{1}{2} \frac{\dot \phi_1^2}{N^2}-V(\phi_1)\right)+\frac{1}{4} \frac{\dot \phi_2^2}{N^2}  + 8\pi G P=0,\nonumber\\
&&\rm
  -\frac{\dot N}{N^2}\left[\frac{\dot m_1}{Nm_1}+\frac{\dot m_2}{Nm_3}+2\frac{\dot \eta}{N\eta}\right]+\frac{\ddot m_1}{N^2m_1}+ \frac{\ddot m_2}{N^2m_2}+\frac{\dot m_1}{Nm_1}\frac{\dot m_2}{Nm_2}  +2\frac{\ddot \eta}{N^2\eta}\label{3,3}\\
&&\mbox{}\rm+\left(\frac{\dot \eta}{N\eta}\right)^2 +3\frac{\dot \eta}{N\eta}\left[\frac{\dot m_1}{Nm_1}+\frac{\dot m_2}{Nm_2} \right] \rm
 +\frac{1}{2}\left(  \frac{1}{2} \frac{\dot \phi_1^2}{N^2}-V(\phi_1)\right)+\frac{1}{4} \frac{\dot \phi_2^2}{N^2}  + 8\pi G P=0, \nonumber
\end{eqnarray}
equations (\ref{0,0}-\ref{3,3}) represent the $\tiny(\begin{tabular}{c} 0\\0 \end{tabular})$, $\tiny( \begin{tabular}{c} 1\\1 \end{tabular})$, $\tiny(\begin{tabular}{c} 2\\2 \end{tabular})$ and the $\tiny(\begin{tabular}{c} 3\\3 \end{tabular})$ components, respectively and the dot (~$\dot{}$~) represents a time derivative. The
corresponding Klein-Gordon (KG) equations are given by
\begin{eqnarray}
\rm \frac{\dot N}{N}\frac{\dot \phi_1^2}{N^2}- \frac{\dot \phi_1
\ddot \phi_1}{N^2}-3\frac{\dot \eta}{\eta} \frac{\dot
\phi_1^2}{N^2}- \dot V&=&\rm 0,\qquad \to \qquad \frac{d}{dt}\, Ln\left(\frac{N}{\eta^3 \dot \phi_1} \right)=\frac{N^2 \dot V}{{\dot \phi_1}^2},\nonumber\\
\rm \frac{\dot N}{N}\frac{\dot \phi_2^2}{N^2}- \frac{\dot \phi_2
\ddot \phi_2}{N^2}-3\frac{\dot \eta}{\eta} \frac{\dot
\phi_2^2}{N^2}&=&\rm 0,\qquad \to \qquad
\frac{d}{dt}\,Ln\left(\frac{N}{\eta^3 \dot \phi_2} \right)=0,
\label{klein-gordon}
\end{eqnarray}
where from the last equation in (\ref{klein-gordon}) it is easy to see that the solution for the scalar field $\rm \phi_2$ (in quadrature form) is given by
\begin{equation}
\rm \Delta\phi_2=\phi_{20}\int \frac{N}{\eta^3}dt,
\end{equation}
with $\rm \phi_{20}$ an integration constant.

It is easy to check that if we performed the subtraction of (\ref{1,1}) from the component (\ref{2,2}), and identifying that
\begin{equation}
\rm \frac{1}{N}\left[ \frac{\dot m_2}{Nm_2} - \frac{\dot m_1}{Nm_1}
\right]^\bullet =\frac{1}{N^2}\left[\frac{\ddot m_2}{m_2} -
\frac{\ddot m_1}{m_1} \right] -\frac{1}{N^2}\left[ \left(\frac{\dot
m_2}{m_2}\right)^2 -\left(\frac{\dot m_1}{m_1}\right)^2\right] +
\frac{\dot N}{N^3}\left[\frac{\dot m_1}{m_1} -\frac{\dot m_2}{m_2}
\right],
\end{equation}
we get
\begin{equation}
\rm \frac{1}{N}\left[ \frac{\dot m_2}{Nm_2} - \frac{\dot m_1}{Nm_1}
\right]^\bullet +3\frac{\dot \eta}{N\eta}\left[\frac{\dot
m_2}{Nm_2}-\frac{\dot m_1}{Nm_1}\right]=0, \label{12}
\end{equation}
where
$[~]^\bullet$ also denotes a time derivative. Defining $\rm R_{21}=(\dot m_2/Nm_2) - (\dot
m_1/Nm_1)$, equation (\ref{12}) can be casted as $ \rm (\dot R_{21}/R_{21})+(3\dot \eta/\eta)=0$, giving a solution
of the form $\rm R_{21}=\ell_{21}/\eta^3$ ($\ell_{12}$ is an integration constant). This procedure can be applied to
components (\ref{1,1})-(\ref{3,3}) arriving at similar expressions, namely, $\rm R_{32}=\ell_{32}/\eta^3$ and $\rm R_{13}=\ell_{13}/\eta^3$,
all three integration constants must satisfy $\rm \ell_{21}+\ell_{32} +\ell_{13}=0$. Now, if take $\rm R_{21}, \rm R_{32}$ and $\rm R_{13}$
together with constraints (\ref{mi}), we obtain the following
\begin{equation}\label{mi's}
\rm \frac{\dot m_2}{Nm_2}=\frac{\ell_2}{\eta^3},\qquad
\rm \frac{\dot m_3}{Nm_3}=\frac{\ell_3}{\eta^3}, \qquad
\rm \frac{\dot m_1}{Nm_1}=\frac{\ell_1}{\eta^3},
\end{equation}
in this las three equations $\ell_2=(\ell_{21}-\ell_{32})/3$, $\rm
\ell_3=(\ell_{32}-\ell_{13})/3$, $\rm
\ell_1=(\ell_{13}-\ell_{21})/3$ and they must satisfy $\sum_{j=1}^3
\ell_j=0$ (for more information on what was discussed above we refer
the reader to Ref. \cite{Socorro:2019vvh}). Now that equations
(\ref{mi's}) have a more manageable form,
 the solutions are straightforward, given by
\begin{equation}
\rm m_i(t)=\delta_i Exp\left[\ell_i \int \frac{Ndt}{\eta^3}\right],
\end{equation}
where $\rm \Pi_{j=1}^3 \delta_j=1$. Setting the gauge $\rm N \to
\eta^3$, the solution becomes
\begin{equation} \rm m_i(t)\to \alpha_i
Exp\left[\ell_i \Delta t\right]. \label{mi-s}
\end{equation}
Unfortunately under this approach we could not find analytical solution for $\eta$, because we need to know the solution for the
scalar field $\phi_1$ (see equation (\ref{0,0})). To be able to reach a solution we are going to resort to the Hamiltonian formalism.
To this end, we employ equation (\ref{first-lagra}) and the line element
(\ref{bianchi}), now the Lagrangian density with the scalar potential field $\rm V(\phi_1)=V_1 e^{-\lambda_1 \phi_1}$ becomes
\begin{equation}\label{lagrafrw-s}\small
\rm {\cal{L}}= \rm
\eta^3\left(\frac{6}{N}\left(\frac{\dot\eta}{\eta}\right)^2-\frac{1}{N}\left[\left(
\frac{\dot m_1}{m_1} \right)^2+\left( \frac{\dot m_2}{m_2}
\right)^2+\left( \frac{\dot m_3}{m_3} \right)^2
\right]-\frac{\dot{\phi_1}^2}{2N}-\frac{\dot{\phi_2}^2}{2N} + N V_1
e^{-\lambda_1 \phi_1}  \right)\,,
\end{equation}
where the momenta are
\begin{equation}
\begin{split}
\rm \Pi_\eta &= \rm 12 \frac{\eta}{N}\dot \eta \\
\rm \Pi_{\phi_1}&=\rm  -\frac{\eta^3}{N}\dot \phi_1\\ \label{momenta-s}
\rm \Pi_{\phi_2}&= \rm -\frac{\eta^3}{N} \dot \phi_2,\\
\rm \Pi_1&= \rm -\frac{2\eta^3}{N}\left( \frac{\dot m_1}{m_1^2}\right),\\
\rm \Pi_2&= \rm -\frac{2\eta^3}{N}\left( \frac{\dot m_2}{m_2^2}\right),\\
\rm \Pi_3&= \rm -\frac{2\eta^3}{N}\left( \frac{\dot m_3}{m_3^2}\right),
\end{split}
\qquad
\begin{split}
\rm\dot \eta&=\rm\frac{N }{12\eta} \Pi_\eta, \\
\rm\dot \phi_1&=\rm -\frac{N}{\eta^3} \Pi_{\phi_1} \\
\rm\dot \phi_2&=\rm - \frac{N}{\eta^3}\Pi_{\phi_2},\\
\rm \dot m_1&=\rm -\frac{N m_1^2\Pi_1}{2\eta^3},\\
\rm \dot m_2&=\rm -\frac{N m_2^2\Pi_2}{2\eta^3},\\
\rm\dot m_3&=\rm -\frac{N m_3^2\Pi_3}{2\eta^3},
\end{split}
\end{equation}
leading to the Hamiltonian density, which takes the form
\begin{equation}\small
\rm {\cal H}= \frac{1}{24\eta}  \Pi_\eta^2- \frac{1}{4\eta^3}m_1^2
\Pi_1^2 - \frac{1}{4\eta^3}m_2^2 \Pi_2^2 - \frac{1}{4\eta^3}m_3^2
\Pi_3^2 -\frac{1}{2\eta^3}
 \Pi_{\phi_1}^2 - \frac{1}{2\eta^3}
\Pi_{\phi_2}^2 -V_1 \eta^3 e^{-\lambda_1\phi_1}  \,.
\label{first-hamibianchi}
\end{equation}
Making the transformation $\rm \Pi_\eta=\partial S/\partial
\eta$, and $\rm \Pi_i=\partial S/\partial m_i$ and choosing
$\rm\eta=e^{u}$ and $\rm m_i=e^{u_i}$, where $\rm P_i=\partial
S/\partial u_i$ and $\rm \pi_u=\partial S/\partial u$, the
Hamiltonian density becomes
\begin{equation}
\rm {\cal H}= \frac{e^{-3u}}{24} \left[ \pi_u^2-6
P_1^2-6P_2^2-6P_3^2- 12
 \Pi_{\phi_1}^2 - 12
\Pi_{\phi_2}^2 -U(u,\phi_1) \right], \label{second-hamibianchi}
\end{equation}
where $\rm U(u,\phi_1)=24V_1e^{6u-\lambda_1\phi_1}$ is the potential function. In the gauge $\rm N=24 e^{3u}$, the Hamilton equations are
\begin{equation}
\begin{split}
\rm \dot u&= \rm 2\pi_u, \\
\rm\dot \pi_u&=\rm 6U,
\end{split}
\qquad
\begin{split}
\dot \phi_1&=\rm-24\Pi_{\phi_1},\\
\dot \phi_2&=\rm-24\Pi_{\phi_2},
\end{split}
\qquad
\begin{split}
\rm\dot \Pi_{\phi_1}&=\rm-\lambda_1 U,\\
\rm\dot \Pi_{\phi_2}&=0,
\end{split}
\qquad
\begin{split}
\rm\dot u_i&= \rm -12 P_i,\\
\rm\dot P_i&=0. \label{hamilton}
\end{split}
\end{equation}
From Hamilton equations (\ref{hamilton}), we can find relations between the scale factor and the scalar fields, which read
\begin{eqnarray}\label{phis}
&& \rm\dot\phi_1 =\rm -24\Pi_{\phi_1}= 4\lambda_1 \pi_u +24p_{\phi_1}=2\lambda_1 \dot u +24 p_{\phi_1},\nonumber\\
&& \rm\dot \phi_2 =\rm -24\Pi_{\phi_2}=24p_{\phi_2},\\
&& \rm\dot u_i=\rm -12p_i,\nonumber
\end{eqnarray}
being $\rm p_{\phi_1}$, $\rm p_{\phi_2}$ and $\rm p_i$ integration constants to be determined by suitable conditions.
The solutions of equations (\ref{phis}) read
\begin{eqnarray}
\rm \Delta \phi_1&=&\rm 2\lambda_1 \Delta u + 24p_{\phi_1} \Delta t, \label{phi}\\
\rm \Delta \phi_2&=& \rm 24p_{\phi_2} \Delta t \,, \label{sigma} \\
\rm\Delta u_i&=&\rm 12p_i \Delta t \label{u_i},\\
\rm  m_i&=&\rm \beta_i e^{-12p_i\Delta t}\label{m_i},
\end{eqnarray}
where the constants $\rm p_i$ must fulfil that $\rm \sum_{i=1}^{3} p_i=0$ and $\rm \sum_{i=1}^3p_i^2=2(p_2^2+p_2p_3+p_3^2)$.
Equations (\ref{phi}-\ref{m_i}) are expressions similar to the solution found by algebraic manipulation to (\ref{klein-gordon}) for the scalar field $\rm \phi_2$ and the Einstein equation (\ref{mi-s}) for the $\rm m_i$ functions. These expressions are indeed general relations since they satisfy the
EKG equations Eqs.(\ref{0,0}-\ref{klein-gordon}).

On the other hand, taking into account the constraint $\rm {\cal
H}=0$, we obtain the temporal dependence for $\rm \pi_u(t)$ which
allows us to construct a master equation:
\begin{equation}
\rm \frac{d \pi_u}{\alpha_1 \pi_u^2 - \alpha_2 \pi_u - \alpha_3}=dt
\,, \label{master-equation}
\end{equation}
where the parameters $\rm \alpha_i$ with $i=1,2,3$, are
\begin{equation}\label{parameter}
\rm \alpha_1=2(3- \lambda_1^2)=2\beta \,,\quad \alpha_2=24\lambda_1
p_{\phi_1} \,, \quad \alpha_3=72\left[p_{\phi_1}^2+ c^2\right]
\,,
\end{equation}
where $\rm c^2=p_{\phi_2}^2+ p_2^2+p_2p_3+p_3^2$.
In the next subsections, we present solutions for three different values of the parameter $\lambda_1$ and also we are able to construct the anisotropic parameters.
\subsection{Case $\alpha_1>0$ and $\lambda_1<\sqrt{3}$}
For this case, we have that the solution for $\rm\pi_u(t)$ is given by
\begin{equation} \rm \pi_u = \frac{1}{4\beta}\left[\alpha_2-\alpha Coth\left(\frac{\alpha}{2}t\right)\right] \,,
\end{equation}
where $\rm \alpha=24\omega_1$ with
$\omega_1=\sqrt{3p_{\phi_1}^2+c^2\beta}$. The solutions of the set
of variables $\rm(u,u_i,\phi_1, \phi_2)$ and $\rm
(\Pi_{\phi_1},\Pi_{\phi_2},P_i)$ are:
\begin{eqnarray}
&&\rm u = u_0 +12\frac{\lambda_1 p_{\phi_1}}{\beta}t+\ln{\left[Csch{\left(12 \omega_1t\right)}\right]}^{1/\beta} \,,\\
&&\rm \phi_1 = p_{\phi_{1_{0}}} +  72\frac{p_{\phi_1}}{\beta} t -Ln{\left[Sinh{\left(12 \omega_1 t\right)}\right]}^{2\lambda_1/\beta} \,, \\
&&\rm \phi_2 = p_{\phi_{2_{0}}} +24p_{\phi_2} t \,, \\
&&\rm    u_i = -12p_i \Delta t\,, \\
&&\rm \Pi_{\phi_1} =-3\frac{p_{\phi_1}}{\beta}+ \frac{\lambda_1 \alpha}{24 \beta}coth \left(\frac{\alpha}{2}t\right) \,, \\
&&\rm \Pi_{\phi_2}= - p_{\pi_2} \,, \\
&&\rm P_i=p_i\,,
\end{eqnarray}
here ($\rm u_0, p_{\phi_{1_{0}}}, p_{\phi_{2_{0}}}, p_i $) are
integration constants. Finally the scale factor $\rm \eta=e^u$ and
the anisotropic parameters take the form
\begin{eqnarray}\label{50}
\rm \eta&=&\rm \eta_0 Exp\left[12\frac{\lambda_1p_{\phi_1}}{\beta}t\right] \, Csch^{\frac{1}{\beta}}
{\left(12\sqrt{3p_{\phi_1}^2+p_{\phi_2}^2\beta} t\right)} \,, \nonumber\\
\rm m_i(t)&=& \rm  \beta_i Exp\left[-12p_i \Delta t \right],
\end{eqnarray}
where $\rm \eta_0=e^{u_0}, \sum_{i=1}^3 p_i=0$ and $\Pi_{i=1}^3\beta_i=1$.
\subsection{Case $\alpha_1<0$ and $\lambda_1>\sqrt{3}$}
In this instance its appropriate to take the relation between the momenta
\begin{equation}
\rm \Pi_{\phi_1}=-\frac{\lambda_1}{6}\pi_u+p_{\phi_1}, \qquad \rm and \qquad
 \Pi_{\phi_2}=-p_{\phi_2}=constant,   \label{momentos-1}
\end{equation}
then the constant $\rm-\alpha_2=24\lambda_1 p_{\phi_1} $, allowing us to obtain the
temporal dependence for $\rm \pi_u(t)$ with which a master equation can be constructed
\begin{equation}
\rm \frac{d \pi_u}{-\alpha_1 \pi_u^2 + \alpha_2 \pi_u - \alpha_3}=dt
\,, \label{master-equation1}
\end{equation}
where we have included the minus sign such the constant $\rm
\alpha_1=2(\lambda_1^2-3)=2\beta>0$. Then, defining $\rm \omega_1^2
=\alpha_2^2 -8\beta \alpha_3=576 \omega_2^2$ with $\omega_2^2=3\rm
p_{\phi_1}^2-(\lambda_1^2-3) c^2$, we can rewrite
(\ref{master-equation1}) as
\begin{equation}
\frac{8\beta \,d \pi_u}{\omega_1^2-\left(4\beta \pi_u-24\lambda_1
\rm p_{\phi_1} \right)^2}=\rm dt,\label{master-integral}
\end{equation}
where the constraint over the parameters $\rm p_{\phi_1}>
c\sqrt{\left(\frac{\lambda_1}{\sqrt{3}}\right)^2-1}$ must be
satisfied. In order to be able to integrate (\ref{master-integral}), as a final step, we resort to the change
of variables $\rm z=4\beta \pi_u-24\lambda_1 p_{\phi_1}$, thus, the
solution for the momenta $\rm \pi_u(t)$ becomes
\begin{equation}
\rm \pi_u=\frac{6\lambda_1p_{\phi_1}}{\beta}+\frac{6\omega_2}{\beta} Tanh\left( 12 \omega_2(t-t_0) \right).
\end{equation}
Using the relations from (\ref{hamilton}) and after some algebra,
the solutions for the set of variables $\rm(u,\phi_1, \phi_2)$ and
$\rm (\Pi_{\phi_1},\Pi_{\phi_2})$ are:
\begin{eqnarray}
&&\rm u= u_0 + \frac{12\lambda_1 p_{\phi_1}}{\beta}(t-t_0) + \frac{1}{\beta}\, Ln\left[Cosh\left(12\omega_2(t-t_0)\right)\right]  \,,\\
&&\rm    u_i = -12p_i \Delta t\,, \\
&&\rm \phi_1 =\phi_{1_{0}} + 72\frac{p_{\phi_1}}{\beta} (t-t_0) +\frac{2\lambda_1}{\beta}\,Ln\left[Cosh\left(12\omega_2(t-t_0)\right) \right] \,, \\
&&\rm \phi_2= \phi_{2_{0}} +24 p_{\phi_2}(t-t_0), \\
&&\rm \Pi_{\phi_1}=-\frac{3p_{\phi_1}}{\beta}-\frac{\lambda_1\omega_2}{\beta} Tanh\left(12\omega_2(t-t_0) \right) \,, \\
&&\rm \Pi_{\phi_2}=-p_{\phi_2}\,,
\end{eqnarray}
where ($\rm u_0,\phi_{1_{0}},\phi_{2_{0}},p_i$) are all
integration constants.  Finally the scale factor becomes
\begin{equation}
\rm \eta=\eta_0\, Exp\left[ \frac{12\lambda_1
p_{\phi_1}}{\beta}(t-t_0) \right]
\,\,Cosh^{\frac{1}{\beta}}\left(12\omega_2(t-t_0)\right),\label{volumen2}
\end{equation}
with $\rm\eta_0=e^{u_0}$ and the anisotropic dimensionless function is
\begin{equation}
\rm m_i(t)=\rm \beta_i Exp\left[ -12p_i \Delta t \right],\label{62}
\end{equation}
we can see that equation (\ref{62}) has the same functional form as before (equation (\ref{50})).
\subsection{Case $\alpha_1=0$ and $\lambda_1^2=3.$}
 For this case the coefficient $\rm \alpha_1=0$ and the master equation to solve is reduced to
\begin{equation}
\rm \int\frac{d \pi_u}{\alpha_2 \pi_u - \alpha_3} = \int dt \,,
\end{equation}
thus the solution for $\rm \pi_u(t)$ can be obtained relatively easily, which read
\begin{equation}\label{Pi-u}
\rm\pi_u(t)=\frac{\alpha_3}{\alpha_2} +p e^{\alpha_2(t-t_0)} \,,
\end{equation}
where $\rm p$ is an integration constant. As before, we can use
relations from Eq.(\ref{hamilton}) and after some manipulation, the
solutions for $\rm(u,\phi_1, \phi_2)$ and $\rm
(\Pi_{\phi_1},\Pi_{\phi_2})$ are:
\begin{eqnarray}
&&\rm u=u_0 +2\sqrt{3} \frac{p_{\phi_1}^2+c^2}{p_{\phi_1}}(t-t_0)+ \frac{\sqrt{3}p}{36p_{\phi_1}} e^{24\sqrt{3}p_{\phi_1}(t-t_0)} \label{u2},\\
&&\rm    u_i = -12p_i \Delta t\,, \\
&&\rm \phi_1=\phi_{1_{0}} + 12\frac{c^2-p_{\phi_1}^2+}{p_{\phi_1}} (t-t_0) + \frac{p}{6} e^{24\sqrt{3}p_{\phi_1}(t-t_0)} \,, \\
&&\rm \phi_2=\phi_{2_{0}}  +24 p_{\phi_2} (t-t_0) \,, \\
&&\rm \Pi_{\phi_1}=\frac{1}{2}\frac{p_{\phi_1}^2-c^2+}{p_{\phi_1}} -\frac{\sqrt{3}p}{6}  e^{24\sqrt{3}p_{\phi_1}(t-t_0)} \,, \\
&&\rm \Pi_{\phi_2}=-p_{\phi_2}, \label{36}
\end{eqnarray}
again ($\rm u_{0},\phi_{1_{0}},\phi_{2_{0}},p_i$) are all
integration constants.  Finally the scale factor $\rm \eta(t)$ for
this case is
\begin{equation}\label{huge-scale}
\rm \eta=\eta_0 Exp\left[2\sqrt{3}
\frac{p_{\phi_1}^2+c^2}{p_{\phi_1}}(t-t_0)\right] \,Exp\left[
\frac{\sqrt{3}p}{36p_{\phi_1}}
e^{24\sqrt{3}p_{\phi_1}(t-t_0)}\right] \,,
\end{equation}
where $\rm \eta_{0}=e^{u_{0}}$, and as before, the anisotropic dimensionless
function $\rm m_i(t)$ is the same as in (\ref{50}).
\subsection{Anisotropic Parameters}
In anisotropic cosmology, the Hubble parameter $\rm H$ is defined in
analogy with the FRW cosmology, that is
\begin{equation}
\rm H=\frac{\dot a}{a}=\frac{\dot \eta }{\eta
}=\frac{1}{3}\left(H_x+H_y+H_z \right),
\end{equation}
where $\rm H_x=\dot a_1/a_1$, $\rm H_y=\dot
a_2/a_2$, and $\rm H_z=\dot a_3/a_3$.

The scalar expansion $\rm \theta$, the shear scalar $\rm \sigma^2$ and
the average anisotropic parameter $\rm \overline A_m$ are defined as
\begin{equation}
\rm \theta=\sum_{i=1}^3 \frac{\dot a_i}{a_i}= 3H, \qquad
\sigma^2=\frac{1}{2} \left(\sum_{i=1}^3 H_i^2 - \frac{1}{3}\theta^2
\right), \qquad \overline A_m = \rm \frac{1}{3} \sum_{i=1}^3 \left(
\frac{H_i-H}{H}\right)^2, \label{aniso-parameter}\end{equation}
respectively.

\begin{figure}[ht!]
\begin{center}
\includegraphics[scale=1]{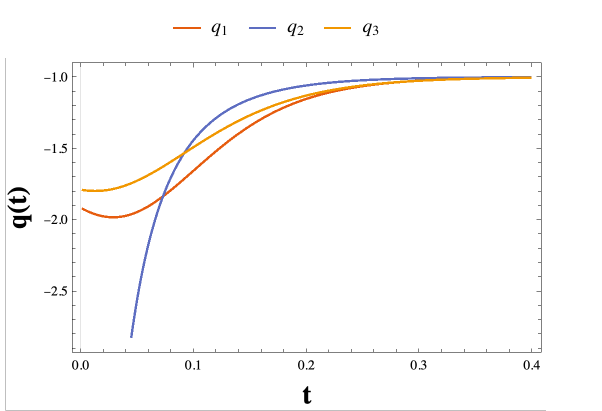}
\caption{Deceleration parameter for the three classical solutions. Here we
have taken $\lambda_1=0.5$, $\lambda_1^{\prime}=2$, $\rm
p_{\phi_1}=0.4$, $\rm p_{\phi_2}=0.2$, $p_2=p_3=0.01$ and
p=0.7.}\label{q-parameters}
\end{center}
\end{figure}

Following Ref. \cite{Tripathy2012}, we consider the volume
deceleration parameter,
\begin{equation}
\rm q(t)=-\frac{v\ddot{v}}{\dot v^2},
\end{equation}
where $\rm v=\eta^3=a_1a_2a_3$ is the (isotropic) volume function of
the Bianchi type I model, and for this case we will have a deceleration parameter for each of the exact solutions given by parameter $\lambda_1$,
 that is
\begin{equation}
\rm q(t)=\left\{
\begin{tabular}{lr}
$\rm -1-\frac{1}{3}\frac{\beta \omega_1^2 }{\left(\lambda_1
p_{\phi_1}Sinh(12\omega_1 \Delta t) -\omega_1 Cosh(12\omega_1 \Delta
t)\right)^2}$, & for $\lambda_1<\sqrt{3}$ \\\\
$\rm -1-\frac{1}{3}\frac{\beta \omega_2^2 }{\left(\lambda_1
p_{\phi_1}Cosh(12\omega_1 \Delta t) +\omega_2 Sinh(12\omega_1 \Delta
t)\right)^2}$, & for $\lambda_1>\sqrt{3}$ \\\\
$-1-\frac{4\sqrt{3}p p_{\phi_1}^3 e^{24\sqrt{3}p_{\phi_1}\Delta
t}}{\left(p p_{\phi_1}e^{24\sqrt{3}p_{\phi_1}\Delta t} +
\sqrt{3}\left[p_{\phi_1}^2+c^2\right] \right)^2}$, & for $\lambda_1=\sqrt{3}$
\end{tabular}
\right.
\end{equation}
these observations indicate that the universe presents a volume
accelerated expansion in the inflationary epoch. Fig.(\ref{q-parameters})
shows the temporal evolution of the deceleration parameter, where $q_1, q_2$
and $q_3$ stand for the solutions $\lambda_1<\sqrt{3}$, $\lambda_1>\sqrt{3}$, $\lambda_1=\sqrt{3}$, respectively.
Using the results for the average scale factor $\eta$ and the
dimensionless anisotropic functions $\rm m_i$, the average
anisotropic parameter is
\begin{equation}\label{Am-parameter}
\rm \overline A_m=\left\{
\begin{tabular}{lr}
$\rm \frac{8}{3} \frac{\beta^2 (\ell_2^2+\ell_2 \ell_3 + \ell_3^2)
Cosh^2(12 \omega_1 \Delta t)}{\left( \omega_1 Cosh(12\omega_1 \Delta t)-\lambda_1 p_{\phi_1} Sinh(12\omega_1 \Delta t)\right)^2}$, & for $\lambda_1<\sqrt{3}$\\\\
$\rm \frac{8}{3}\frac{\beta^2\left(\ell_2^2+\ell_2 \ell_3 +
\ell_3^2\right)Cosh^2(12\omega_2 \Delta
t)}{\left(\omega_2\,Sinh(12\omega_2 \Delta t) +\lambda_1 p_{\phi_1} Cosh(12\omega_2 \Delta t)\right)^2}$ , & for $\lambda_1>\sqrt{3}$\\\\
$\frac{96p_{\phi_1}^2\left(\ell_2^2+\ell_2\ell_3+\ell_3^2
\right)}{\left(p p_{\phi_1}e^{24\sqrt{3}p_{\phi_1}\Delta t} +
\sqrt{3}\left[p_{\phi_1}^2+c^2\right] \right)^2}$. & for
$\lambda_1=\sqrt{3}$
\end{tabular}
\right.
\end{equation}
The other two parameters acquire the form
\begin{equation}
\rm \theta = \left\{ \begin{tabular}{lr}
$\rm \frac{36}{\beta} \left(\lambda_1 p_{\phi_1}-\omega_1 Ctgh(12\omega_1 \Delta t) \right)$, & for $\lambda_1<\sqrt{3}$, \\\\
$\rm \frac{36}{\beta}\left[\lambda_1 p_{\phi_1}+\omega_2 Tanh(12\omega_2 \Delta t) \right]$, & for $\lambda_1>\sqrt{3}$ \\\\
 $\rm 6\left(p\, e^{24\sqrt{3}p_{\phi_1}\Delta
t}+\sqrt{3}\frac{p_{\phi_1}^2 +c^2}{p_{\phi_1}} \right)$, & for
$\lambda_1=\sqrt{3}$
\end{tabular}
\right. \label{expansion}
\end{equation}

\begin{equation}
\rm \sigma^2= \left\{ \begin{tabular}{lr} $\rm
648\frac{\frac{8}{9}\ell^2 Sinh^2(\gamma)-\left(\omega_1
Cosh(\gamma)-
\lambda_1 p_{\phi_1}Sinh(\gamma)\right)^2}{\beta^2 Sinh^2(\gamma)}$, & for $\lambda_1<\sqrt{3}$\\\\
 $\rm 648\frac{\frac{8}{9}\ell^2Cosh^2(\gamma) -\left(\omega_2
Sinh(\gamma)+
\lambda_1 p_{\phi_1}Cosh(\gamma)\right)^2}{\beta^2\, Cosh^2(\gamma)} $, & for $\lambda_1> \sqrt{3}$ \\\\
 $\rm 18\left\{32\ell^2 -\left[ \frac{p\,p_{\phi_1} e^{24\sqrt{3}p_{\phi_1}\Delta
t}+\sqrt{3}\left(p_{\phi_1}^2 +c^2 \right)} {p_{\phi_1}}\right]^2
\right\}$, & for $\lambda_1=\sqrt{3}$
 \end{tabular}
 \right.  \label{shear}
\end{equation}
where we have define $\gamma=12\omega_1 \Delta t$ and
$\ell^2=\ell_2^2+\ell_2 \ell_3 + \ell_3^2$ strictly for format
reasons. In Ref. \cite{Pradhan:2010dm} and references therein, the
authors pin down that the red-shift studies place the limit
$\sigma/\theta\leq 0.3$ on the ratio of shear $\sigma$ to Hubble
constant $H$ in the neighborhood of our Galaxy today in order to
have a sufficiently isotropic cosmological model, in this regard we
obtain
\begin{equation}
\rm \frac{\sigma^2}{\theta^2}= \left\{
\begin{tabular}{lr} $\rm-\frac{1}{2}+\frac{4}{9}\frac{\beta^2 \left(\ell_2^2+\ell_2 \ell_3 +
\ell_3^2 \right) Sinh^2(12\omega_1 \Delta t)}{\left(\omega_1 Cosh(12\omega_1 \Delta t) -\lambda_1p_{\phi_1}\,Sinh(12\omega_1\Delta t)\right)^2 }$, &for $\lambda_1<\sqrt{3}$ \\\\
$\rm-\frac{1}{2}+\frac{4}{9}\frac{\beta^2(\ell_2^2+\ell_2\ell_3+\ell_3^2)\,Cosh^2(12\omega_2 \Delta) t}{\left(\lambda_1
p_{\phi_1}\,Cosh(12\omega_2 \Delta t)+\omega_2 Sinh(12\omega_2\Delta t)\right)^2}$, & for $\lambda_1>\sqrt{3}$  \\\\
$\rm
-\frac{1}{2}+\frac{16p_{\phi_1}^2\left(\ell_2^2+\ell_2\ell_3+\ell_3^2\right)}{\left[
p\,p_{\phi_1}e^{24\sqrt{3}p_{\phi_1}\Delta
t}+\sqrt{3}\left(p_{\phi_1}^2 +c^2\right)\right]^2} $, & for
$\lambda_1=\sqrt{3}$
\end{tabular} \right. \label{reason}
\end{equation}
from Eqs. (\ref{Am-parameter}) and (\ref{reason}) we can constraint the
average anisotropic parameter $\rm \overline A_m$ to the following
value for both $\lambda_1>\sqrt{3}$ and $\lambda_1=\sqrt{3}$:
$\rm\overline A_m\leq3.54$, signaling that the anisotropic phase still continues.
\section{Second model: chiral anisotropic model}\label{second-model}
Now we turn our attention to the second model to be considered. In this case, the action for such a universe is given by
\begin{equation}
\rm {\cal L}=\sqrt{-g} \left( R-\frac{1}{2}g^{\mu\nu}
m^{ab}\nabla_\mu \phi_a \nabla_\nu \phi_b  + V(\phi_1,\phi_2)\right)
\,, \label{lagra}
\end{equation}
where $\rm R$ is the Ricci scalar, $\rm V(\phi_1,\phi_2)=V_1 e^{-\lambda_1 \phi_1} + V_2
e^{-\lambda_2 \phi_2}$ is the
corresponding scalar field potential, and $\rm m^{ab}$ is a
 $2 \times 2$ constant matrix and $\rm m^{12}=m^{21}$. The EKG equations are obtained varying Eq.(\ref{lagra}) with respect to the metric
  and the scalar fields, resulting in
\begin{equation}
\rm G_{\alpha \beta}=\rm -\frac{1}{2}m^{ab} \left(\nabla_\alpha
\phi_a \nabla_\beta \phi_b -\frac{1}{2}g_{\alpha \beta} g^{\mu \nu}
\nabla_\mu \phi_a \nabla_\nu \phi_b \right) +\frac{1}{2}g_{\alpha
\beta} \, V(\phi_1,\phi_2), \label{mono}
\end{equation}
\begin{equation}
\rm m^{ab}\Box \phi_b -\frac{\partial V}{\partial \phi_a} =\rm
m^{ab}g^{\mu\nu} {\phi_b}_{,\mu\nu} - m^{ab}g^{\alpha \beta}
\Gamma^\nu_{\alpha \beta} \nabla_\nu \phi_b - \frac{\partial
V}{\partial \phi_a}=\rm 0, \,\qquad a,b=1,2.\label{ekg-phi}
\end{equation}
Consequently the Klein-Gordon equations are
\begin{eqnarray}
\rm m^{11}{\phi_1^{\prime \prime}}{\phi_1^\prime}+
m^{12}{\phi_2^{\prime \prime}}{\phi_1^\prime}
+3\frac{\eta^\prime}{\eta} \left(m^{11}{\phi_1^\prime}^2 + m^{12}
\phi_1^\prime \phi_2^\prime \right) +\left(\dot V\right)_{\phi_2}
&=&\rm 0 \,,  \label{ein2}\\
\rm m^{22}{\phi_2^{\prime \prime}}{\phi_2^\prime}+
m^{12}{\phi_1^{\prime \prime}}{\phi_2^\prime} +3
\frac{\eta^\prime}{\eta}\left(m^{22}{\phi_2^\prime}^2 + m^{12}
\phi_1^\prime \phi_2^\prime \right)+\left(\dot V\right)_{\phi_1} &=&
\rm0, \, \label{ein3}
\end{eqnarray}
here $\prime=d/d\tau$, $\rm d\tau=Ndt$ and $(\rm\dot V)_{\phi_i}$ means that the derivative is calculated maintaining $\phi_i$ constant
(with \rm i=1,2). An equivalent form to write equations (\ref{ein2}) and (\ref{ein3}) is
\begin{eqnarray}
\rm  m^{11}\dot \phi_1\frac{d}{dt}Ln\left(\frac{N}{\eta^3 \dot
\phi_1}\right)+m^{12}\dot \phi_2\frac{d}{dt}Ln\left(\frac{N}{\eta^3
\dot \phi_2}\right)&=&\rm \frac{N^2 \left(\dot
V\right)_{\phi_2}}{\dot \phi_1},\label{ein21}\\
\rm  m^{12}\dot \phi_1\frac{d}{dt}Ln\left(\frac{N}{\eta^3 \dot
\phi_1}\right)+m^{22}\dot \phi_2\frac{d}{dt}Ln\left(\frac{N}{\eta^3
\dot \phi_2}\right)&=&\rm \frac{N^2 \left(\dot
V\right)_{\phi_1}}{\dot \phi_2}. \label{ein31}
\end{eqnarray}
Taking the metric (\ref{biachi_I_misner}) and pluging it into
(\ref{lagra}), the Lagrangian density becomes
\begin{eqnarray}\label{lagra-mixto}
\rm {\cal{L}}&=& \rm
\eta^3\left(\frac{6}{N}\left(\frac{\dot\eta}{\eta}\right)^2-\frac{1}{N}\left[\left(
\frac{\dot m_1}{m_1} \right)^2+\left( \frac{\dot m_2}{m_2}
\right)^2+\left( \frac{\dot m_3}{m_3} \right)^2
\right]\right.\nonumber\\
&&\rm \left.
-m^{11}\frac{\dot{\phi_1}^2}{2N}-m^{22}\frac{\dot{\phi_2}^2}{2N}-m^{12}\frac{\dot
\phi_1 \dot \phi_2}{N} + N\left[ V_1 e^{-\lambda_1 \phi_1} + V_2
e^{-\lambda_2 \phi_2}\right] \right)\,,
\end{eqnarray}
and the momenta are
\begin{equation}
\begin{split}
\rm \Pi_\eta &= \rm 12 \frac{\eta}{N}\dot \eta,\\
\rm \Pi_{\phi_1}&=\rm -\frac{\eta^3}{N}\left(m^{11}\dot \phi_1+m^{12}\dot \phi_2\right),\\
\rm \Pi_{\phi_2}&= \rm -\frac{\eta^3}{N}\left(m^{22} \dot\phi_2+m^{12}\dot \phi_1\right),\\
\rm \Pi_1&= \rm -\frac{2\eta^3}{N}\left( \frac{\dot m_1}{m_1^2}\right),\\
\rm \Pi_2&= \rm -\frac{2\eta^3}{N}\left( \frac{\dot m_2}{m_2^2}\right),\\
\rm \Pi_3&= \rm -\frac{2\eta^3}{N}\left( \frac{\dot m_3}{m_3^2}\right),
\end{split}
\qquad
\begin{split}
\dot \eta&=\frac{N }{12\eta} \Pi_\eta, \\
\dot \phi_1&=\rm  \frac{N}{\eta^3\triangle}\left(-m^{22} \Pi_{\phi_1}+m^{12} \Pi_{\phi_2} \right),\label{momenta-m} \\
\dot\phi_2&=\frac{N}{ \eta^3 \Delta} \left(m^{12}\Pi_{\phi_1}-m^{11}\Pi_{\phi_2} \right),\\
\dot m_1&=-\frac{N m_1^2\Pi_1}{2\eta^3},\\
\dot m_2&=-\frac{N m_2^2\Pi_2}{2\eta^3},\\
\dot m_3&=-\frac{Nm_3^2\Pi_3}{2\eta^3},
\end{split}
\end{equation}
where $\rm \triangle =m^{11}m^{22}-(m^{12})^2$.  Writing
(\ref{lagra-mixto}) in a canonical form, {\it i.e.}
 $\mathcal L_{can}=\Pi_q\dot q-N\mathcal H$, we can perform the variation of this canonical Lagrangian with respect to the
 lapse function $N$, $\delta\mathcal L_{can}/\delta N=0$, resulting in the constraint $\mathcal H=0$, and making the same transformation as
  in (\ref{second-hamibianchi}), the Hamiltonian density results in
\begin{align}
\rm {\cal H}=\rm &\frac{e^{-3u}}{24} \biggl[\rm \Pi_u^2-6P_1^2-6P_2^2-6P_3^2-\frac{12 m^{22}}{\triangle}\Pi_{\phi_1}^2
-\frac{12 m^{11}}{\triangle}\Pi_{\phi_2}^2+\frac{24m^{12}}{\triangle} \Pi_{\phi_1}\Pi_{\phi_2}\nonumber\\
& \rm-24V_1e^{-\lambda_1\phi_1+6u}-24V_2 e^{-\lambda_2\phi_2+6u}\biggr].\label{hamibi}
\end{align}
Proposing the following canonical transformation on the variables
$\rm(\eta,\phi_1,\phi_2,u_i)\leftrightarrow (\xi_1,\xi_2,\xi_3,u_i)$
\begin{equation}\label{trans_2}
\begin{split}
\rm \xi_1&=\rm-6u+\lambda_1 \phi_1,\\
\xi_2&= \rm-6 u+\lambda_2 \phi_2,\\
 \xi_3&=-\rm4u + \frac{\lambda_1}{6}\phi_1 + \frac{\lambda_2}{6}
 \phi_2,\\
\rm u_i &= \rm u_i,
\end{split}
\quad\longleftrightarrow\quad
\begin{split}
\rm u&=\rm\frac{\xi_1 + \xi_2- 6 \xi_3}{12},\\
\rm \phi_1&= \rm \frac{3\xi_1 + \xi_2-6\xi_3}{2\lambda_1},\\
\rm \phi_2 &= \rm \frac{\xi_1+3\xi_2-6\xi_3}{2\lambda_2},
\end{split}
\end{equation}
and setting the gauge $\rm N=24e^{3u}$, allows us to find a new set
of conjugate momenta $\rm (\pi_1,\pi_2,\pi_3)$
\begin{align} \label{new-moment}
\rm \Pi_u &= \rm -6 \pi_1 -6 \pi_2-4 \pi_3, \nonumber\\
\rm \Pi_{\phi_1} &= \rm \lambda_1 \pi_1 +\frac{\lambda_1}{6}\pi_3,\\
\rm \Pi_{\phi_2} &= \rm \lambda_2 \pi_2 +\frac{\lambda_2}{6}\pi_3, \nonumber
\end{align}
which finally leads us to the Hamiltonian density
\begin{align}
\rm {\cal H} =& \rm  12 \left(3-\frac{\lambda_1^2
m^{22}}{\triangle}\right)\pi_1^2+12\left(3-\frac{\lambda_{2}^2\*m^{11}}{\triangle}\right)\pi_2^2\nonumber\\
&+\left(16 + \frac{- \lambda_1^2 m^{22}+ 2\lambda_1\lambda_2 m^{12}-\lambda_2^2m^{11}}{3\triangle}\right)\pi_3^2\nonumber\\
&\rm + 12\left[   \left(4+ \frac{ \lambda_1 \lambda_2m^{12}-\lambda_1^2 m^{22}}{3\triangle}\right)\pi_1+  \left(4+\frac{\lambda_1 \lambda_{2}m^{12}-\lambda_2^2m^{11}}{3\triangle}\right)\pi_2\right]\pi_3 \nonumber\\
&\rm -6P_1^2-6P_2^2-6P_3^2+24\left(3+\frac{\lambda_1\lambda_2m^{12}}{\triangle}\right)\pi_1\pi_2- 24\left(V_1 e^{-\xi_1}+V_2
e^{-\xi_2}\right),\label{hamiltonian}
\end{align}
the parameter $\rm \triangle$ is the same that was defined after
Eqns. (\ref{momenta-m}). The form that the Hamiltonian density
(\ref{hamiltonian}) acquires after applying the transformation
(\ref{trans_2}) into Eq. (\ref{hamibi}) will, in the end, allows us to
obtain the solutions for this model. First, let's compute Hamilton's
equations, which read
\begin{eqnarray}
\rm \dot \xi_1&=& \rm 24
\left(3-\frac{\lambda_1^2m^{22}}{\triangle}\right)\pi_1
+24\left(3+\frac{\lambda_1\lambda_2m_{12}}{\triangle}\right)\pi_2+
12 \left(4+ \frac{\lambda_1 \lambda_2 m^{12}-\lambda_1^2
m^{22}}{3\triangle}\right)\pi_3,\nonumber\\
\rm \dot \xi_2&=& \rm
24\left(3-\frac{\lambda_{2}^2\*m^{11}}{\triangle}\right)\pi_2
+24\left(3+\frac{\lambda_1\lambda_2m_{12}}{\triangle}\right)\pi_1+
12\left(4+\frac{\lambda_1\lambda_2
m^{12}-\lambda_2^2m^{11}}{3\triangle}\right)\pi_3, \nonumber\\
\rm \dot \xi_3 &=&\rm 12\left[\left(4+ \frac{ \lambda_1 \lambda_2m^{12}-\lambda_1^2 m^{22}}{3\triangle}\right)\pi_1+  \left(4
+\frac{\lambda_1 \lambda_{2}m^{12}-\lambda_2^2m^{11}}{3\triangle}\right)\pi_2\right] \nonumber\\
&&+2 \left(16 + \frac{-\lambda_1^2 m^{22}+2\lambda_1 \lambda_2m^{12}-\lambda_2^2m^{11}}{3\triangle}\right)\pi_3,\label{ecs_mov_2}\\
 \rm  \dot \pi_1&=&\rm  -24 V_1e^{-\xi_1}, \qquad \rm \dot P_i=0,\nonumber\\
 \rm \dot \pi_2&=&\rm -24V_2e^{-\xi_2},\rm \qquad\dot u_i=-12P_i,\nonumber \\
 \rm \dot \pi_3&=&0,\nonumber
\end{eqnarray}
from this last set of equations is straightforward to see that $\rm\pi_3=p_3$ and  $\rm P_i=n_i$ are constants and the solutions
to $\rm u_i=u_{i_{0}}-12n_i \Delta t$. Taking the time derivative of the first equation in (\ref{ecs_mov_2}), we obtain
\begin{equation}\rm
\rm \ddot \xi_1= \rm -576V_1 \left(3-\frac{\lambda_1^2
m^{22}}{\triangle}\right) e^{-\xi_1}
-576V_2\left(3+\frac{\lambda_1\lambda_2m^{12}}{\triangle}\right)
e^{-\xi_2}. \label{first}
\end{equation}
The main purpose of introducing the transformation (\ref{trans_2})
was to be able to separate the set of equations arising from the
Hamiltonian density (\ref{hamiltonian}). To reach a solution to
our problem we set to zero the coefficient that is multiplying the
mixed momenta term in (\ref{hamiltonian}), which sets the following
constraint on the matrix element $\rm m^{12}$
\begin{equation}\label{m12}
\rm m^{12}=\frac{\lambda_1 \lambda_2}{6}\left(1 \pm \sqrt{1+ 36
\frac{m^{11} m^{22}}{\lambda_1^2 \lambda_2^2}}\right),
\end{equation}
the latter implies that the second term in the square root of (\ref{m12})
is a real number, say $\ell=\rm 36(m^{11} m^{22}/\lambda_1^2
\lambda_2^2)$
 $\in\mathbb{R}^+$, giving the same weight to the matrix elements $\rm m^{11}$ and $\rm m^{22}$,
 whose values are $\rm m^{11}=\frac{1}{6}\sqrt{\ell}\lambda_1^2$ and $\rm m^{22}=\frac{1}{6}\sqrt{\ell} \lambda_2^2$. Here, we are going
 to distinguish two possible scenarios for $\rm m^{12}$ as: $\rm m^{12}_+=\frac{1}{6}\lambda_1 \lambda_2\left(1 + \sqrt{1+\ell}\right)>0$
 and $\rm m^{12}_-= -\frac{1}{6} \lambda_1 \lambda_2 \left(\sqrt{1+\ell}-1\right)<0$. This two choices of $\rm m^{12}$ enables us to have a quintom
 like case and quintessence like case, respectively. With these two possible values for the matrix element $\rm m^{12}$ we can see that
$\triangle_+=-\frac{1}{18}\lambda_1^2 \lambda_2^2
\left(1+\sqrt{1+\ell}\right) <0$ for $\rm m^{12}_+$ and $\triangle_-
=\frac{1}{18}\lambda_1^2 \lambda_2^2 \left(\sqrt{1+\ell}-1\right)>0$
for $\rm m^{12}_-$.


\subsection{Quintom like case}
We begin by analyzing the quintom like case, for which the matrix
element $\rm m^{12}_-= -\frac{1}{6}(\sqrt{1+\ell}-1) \lambda_1
\lambda_2$,  the Hamiltonian density is rewritten as,
\begin{align}
\rm {\cal H}=&-\frac{\pi_1^2}{\mu_{_\ell}}-\frac{\pi_2^2}{\mu_{_\ell}} +\left(48-\frac{1}{3c_{_\ell}}
\right)\left(\pi_1+\pi_2 \right)\pi_3+\left(16-\frac{1}{18c_{_\ell}}\right)\pi_3^2-6\left(\rm P_1^2+P_2^2+P_3^2\right)\nonumber\\
&\rm -24V_1e^{\xi_1}-24V_2e^{-\xi_2}, \label{hami-bianchi-q}
\end{align}
also we have defined the parameters $\rm
\mu_{_\ell}=\sqrt{\ell}/36\left(1+\sqrt{1+\ell}-\sqrt{\ell}\right)$
and $\rm c_{_\ell}=\sqrt{\ell}/{36\left[\left(1+\sqrt{1+\ell}\right)+\sqrt{\ell}\right]}$. Thus, Hamilton equations for the new
simplified coordinates $\rm\xi_i$ are
\begin{eqnarray}\label{new_ecs_mov_2}
\rm \dot \xi_1&=& \rm -\frac{2\pi_1}{\mu_{_\ell}}
+\left(48-\frac{1}{3c_{_\ell}} \right)\pi_3, \nonumber\\
\rm \dot \xi_2&=& \rm -\frac{2\pi_2}{\mu_{_\ell}}
+\left(48-\frac{1}{3c_{_\ell}} \right)\pi_3, \\
\rm \dot \xi_3&=& \rm +\left(48-\frac{1}{3c_{_\ell}}
\right)\left(\pi_1+\pi_2 \right) +
2\left(16-\frac{1}{18c_{_\ell}}\right)\pi_3, \nonumber
\end{eqnarray}
the equations for $\rm \dot \pi_i$ remain the same as in Eqs.
(\ref{ecs_mov_2}). Taking the derivative of the first equation of
(\ref{new_ecs_mov_2}) yields
\begin{equation}
\rm \ddot \xi_1= \frac{48V_1}{\mu_{_\ell}}  e^{-\xi_1},
\end{equation}
which has a solution of the form
\begin{equation}
\rm e^{-\xi_1}=\frac{\mu_{_\ell} r_1^2}{24 V_1} \, Sech^2\left(r_1
t-q_1\right). \label{solucion-xi1}
\end{equation}
From (\ref{new_ecs_mov_2}) we can see that $\dot\xi_2$ has the same
functional structure as $\dot\xi_1$, therefore its solution will be
of the same form as (\ref{solucion-xi1}), so we have
\begin{equation}
\rm e^{-\xi_2}=\frac{\mu_{_\ell} r_2^2}{24 V_2} \, Sech^2\left(r_2
t-q_2\right), \label{solucion-xi2}
\end{equation}
where $\rm r_i$ and $\rm q_i$ (with $\rm i=1,2$) are integration
constants, both at Eq. (\ref{solucion-xi1}) and Eq. (\ref{solucion-xi2}).
Reinserting these solutions into Hamilton's equations for the momenta,
we obtain
\begin{eqnarray}
\rm \pi_1 &=& \rm \alpha_1 - \mu_{_\ell} \,r_1 \, Tanh\left(r_1t-q_1\right), \label{solucion-p1}\\
\rm \pi_2 &=& \rm \alpha_2 - \mu_{_\ell}\,r_2 \, Tanh\left(r_2
t-q_2\right). \label{solucion-p2}
\end{eqnarray}
With (\ref{solucion-p1}) and (\ref{solucion-p2}), it can be easily
checked that the Hamiltonian is identically null when
\begin{equation}
\rm \alpha_1=\alpha_2=\frac{72\mu_{_\ell}-1}{6} p_3, \qquad
p_3^2=\frac{\mu_\ell(r_1^2+ r_2^2)+6n^2}{4(72\mu_{_\ell}+1)},
\end{equation}
where $\rm n^2$ belongs to the contribution on the anisotropic
functions and is given by $\rm n^2=n_1^2+n_2^2+n_3^2$.  Now we are in position write the solutions for the $\rm
\xi_i$ coordinates, which read
\begin{align}
\rm \xi_1 =& \rm \beta_1 +Ln\left[ Cosh^2\left(r_1 t -q_1 \right)\right],  \label{xi1}\\
\xi_2 =& \rm \beta_2+Ln\left[ Cosh^2\left(r_2 t -q_2 \right)\right], \label{xi2}\\
\rm \xi_3=&\rm \beta_3 + p_3\left[16 \left(1+72\mu_{_\ell}
\right)-8\frac{\mu_{_\ell}}{c_{_{\ell}}}\right]\Delta t- \left(48
-\frac{1}{3c_{_\ell}}\right)\mu_{_\ell}\times \nonumber\\
&\rm Ln\,\left[Cosh\left(r_1t-q_1 \right)\,Cosh\left(r_2 t-q_2 \right)
\right],
\end{align}
here the $\rm \beta_i$, (with $\rm i=1,2,3$),
terms are constants coming from integration. Applying the inverse
canonical transformation we obtain the solutions in the original
variables $\rm (\eta, \phi_1, \phi_2)$ as
\begin{equation}\label{sols_quintom}
{\small
\begin{split}
\rm \eta &= \rm\eta_0 +\frac{1}{12} Ln\left[ Cosh^2 \left(r_1 t -q_1
\right) Cosh^2\left(r_2 t -q_2 \right)\right]
-\frac{1}{2}p_3\left[16 \left(1+72\mu_{_\ell}
\right)-8\frac{\mu_{_\ell}}{c_{_{\ell}}}\right]  \Delta t \\
& \rm + \frac{1}{2}\mu_\ell\,\left(48 -\frac{1}{3c_{_\ell}}\right)
Ln\,\left[Cosh\left(r_1t-q_1 \right)\,Cosh\left(r_2 t-q_2 \right) \right],\\
\rm \phi_1 &=\rm \phi_{10}+\frac{1}{2\lambda_1}
Ln\left[Cosh^6\left(r_1t-q_1\right)Cosh^2\left(r_2 t
-q_2\right)\right] - \frac{3}{\lambda_1}p_3\left[16
\left(1+72\mu_{_\ell}
\right)-8\frac{\mu_{_\ell}}{c_{_{\ell}}}\right]\Delta t
\\&\rm +\frac{3}{\lambda_1}\mu_\ell
\left(48-\frac{1}{3c_{_\ell}}\right) Ln\,\left[Cosh
\left(r_1t-q_1\right)\,Cosh\left(r_2 t-q_2 \right) \right]
 ,\\
\rm \phi_2 &= \rm
\phi_{20}+\frac{1}{2\lambda_2}Ln\left[Cosh^2\left(r_1t-q_1\right)
Cosh^6\left(r_2 t -q_2\right)\right] -
\frac{3}{\lambda_2}p_3\left[16 \left(1+72\mu_{_\ell}
\right)-8\frac{\mu_{_\ell}}{c_{_{\ell}}}\right]\Delta t\\
&+ \frac{3}{\lambda_2}\mu_\ell\left(48-\frac{1}{3c_{_\ell}}\right) \rm Ln\,\left[Cosh
\left(r_1t-q_1 \right)\,Cosh\left(r_2 t-q_2 \right) \right] ,
\end{split}}
\end{equation}
where $\eta_0, \phi_{10}$ and $\phi_{20}$ are given in terms of the
$\beta_i$ constants as
\begin{equation}
\eta_0=\frac{\beta_1+\beta_2-6\beta_3}{12}, \quad
\phi_{10}=\frac{3\beta_1+\beta_2-6\beta_3}{2\lambda_1},\quad
\phi_{20}=\frac{\beta_1+3\beta_2-6\beta_3}{2\lambda_2}.\label{phi2}
\end{equation}

\subsection{Quintessence like case }

Now we turn our attention to the quintessence like case, for which
the matrix element $\rm m^{12}_+= \frac{1}{6}\left(1 +
\sqrt{1+\ell}\right) \lambda_1 \lambda_2$, then the Hamiltonian
density describing this quintessence model is rewritten as
\begin{align}
\rm {\cal H}&= \frac{\pi_1^2}{\nu_{_\ell}} +
\frac{\pi_2^2}{\nu_{_\ell}} +\left(48-\frac{1}{3c_{_\ell}}
\right)\left(\pi_1+\pi_2 \right)
\pi_3+\left(16-\frac{1}{18c_{_\ell}}\right)\pi_3^2-6\left(\rm P_1^2+P_2^2+P_3^2\right)\nonumber \\
&\rm -24V_1e^{-\xi_1}-24V_2 e^{-\xi_2}, \label{hami-bianchi-qi}
\end{align}
here we define the parameter $\rm\nu_{_\ell}=\sqrt{\ell}/{36\left(\sqrt{1+\ell}+\sqrt{\ell}-1\right)}$
and $\rm c_{_\ell}=\sqrt{\ell}/{36\left(+\sqrt{\ell}+1-\sqrt{1+\ell}\right)}$.

From (\ref{hami-bianchi-qi}) we can calculate Hamilton equations for
the phase space spanned by $\rm (\xi_i, \pi_i)$, given by
\begin{eqnarray}
\rm \dot \xi_1&=& \rm \frac{2\pi_1}{\nu_{_\ell}}
+\left(48-\frac{1}{3c_{_\ell}} \right)\pi_3, \nonumber \\
\rm \dot \xi_2&=& \rm \frac{2\pi_2}{\nu_{_\ell}}
+\left(48-\frac{1}{3c_{_\ell}} \right)\pi_3, \label{xi_quintessence}\\
\rm \dot \xi_3&=& \rm \left(48-\frac{1}{3c_{_\ell}}
\right)\left(\pi_1+\pi_2 \right) +
2\left(16-\frac{1}{18c_{_\ell}}\right)\pi_3,\nonumber\\
\rm P_i &=& \rm n_i=constant, \nonumber
\end{eqnarray}
as in the quintom case $\rm \dot \pi_i$ remain the same as in Eq.
(\ref{ecs_mov_2}). Proceeding in a similar way as in the previous
case, we take the derivative of the first equation in
(\ref{xi_quintessence}), obtaining
\begin{equation}
\rm \ddot \xi_1= -\frac{48V_1}{\nu_{_\ell}}  e^{-\xi_1},
\end{equation}
which the corresponding solution is
\begin{equation}
\rm e^{-\xi_1}=\frac{\nu_{_\ell} r_1^2}{24 V_1} \,
Csch^2\left(r_1t-q_1\right). \label{solution-xi1-1}
\end{equation}
Also in this quintessence like setting, the $\dot\xi_2$ functional
form is the same as $\dot\xi_1$, indicating that the solution is of
the same type as (\ref{solution-xi1-1}), that is
\begin{equation}
\rm e^{-\xi_2}=\frac{\nu_{_\ell} r_2^2}{24 V_2} \, Csch^2\left(r_2
t-q_2\right), \label{solution-xi2-1}
\end{equation}
in Eq. (\ref{solution-xi1-1}) and Eq. (\ref{solution-xi2-1}) the $\rm r_i$
and $\rm q_i$ (with $\rm i=1,2$) are constants coming from
integration. With Eq. (\ref{solution-xi1-1}) and Eq. (\ref{solution-xi2-1})
at hand, we can reinsert them into Hamilton equations for the
momenta, giving
\begin{eqnarray}
\rm \pi_1 &=& \rm -a_1 + \nu_{_\ell} \,r_1 \, Coth\left(r_1 t-q_1
\right), \label{solution-p1-1}\\
\rm \pi_2 &=& \rm -a_2 + \nu_{_\ell}\,r_2 \, Coth\left(r_2 t-q_2
\right), \label{solution-p2-1}
\end{eqnarray}
where it can be easily verify that with this last two equations, the Hamiltonian is identically zero
when
\begin{equation}
\rm a_1=a_2=\frac{72\nu_{_\ell}+1}{6} p_3, \qquad
p_3^2=\frac{\nu_{_\ell}(r_1^2+ r_2^2)+6n^2}{4(72\nu_{_\ell}-1)},
\end{equation}
where $\rm n^2=n_1^2+n_2^2+n_3^2$. So, the solutions for the $\rm \xi_i$ coordinates become
\begin{align}
\rm \xi_1 &= \rm \beta_1 +Ln\left[ Sinh^2
\left(r_1 t -q_1 \right)\right],  \label{xi1-1}\\
\xi_2 &= \rm \beta_2+Ln\left[ Sinh^2
\left(r_2 t -q_2 \right)\right], \label{xi2-1}\\
\rm \xi_3&= \rm \beta_3 -p_3
\left[16\left(72\nu_{_\ell}-1\right)-8\frac{\nu_{_\ell}}{c_{_\ell}}\right]
\Delta t+ \left(48 -\frac{1}{3c_{_\ell}}\right)\,\nu_{_\ell}\times\nonumber\\
&\rm Ln\,\left[Sinh \left(r_1t-q_1 \right)\,Sinh\left(r_2 t-q_2 \right)
\right],
\end{align}
 where $\rm \beta_i$ are integration
constants (with $\rm i=1,2,3$). After applying the inverse canonical transformation we
get the solutions in terms of the original variables $\rm (\Omega,
\phi_1, \phi_2)$ as
\begin{equation}\label{sols_quintessence}
\begin{split}
\rm \eta &= \rm  \eta_0 +\frac{1}{12} Ln\left[ Sinh^2 \left(r_1t
-q_1 \right) Sinh^2\left(r_2 t -q_2 \right)\right]
+\frac{1}{2}p_3\left[16\left(72\nu_\ell-1\right)-8\frac{\nu_{_\ell}}{c_{_\ell}}\right]\Delta t \\
&\rm - \frac{1}{2}\left(48 -\frac{1}{3c_{_\ell}}\right)\,\nu_\ell
Ln\,\left[Sinh \left(r_1t-q_1 \right)\,Sinh\left(r_2 t-q_2 \right) \right], \\
 \rm \phi_1 &= \rm \phi_{10}+\frac{1}{2\lambda_1}\biggl[ Ln\left[Sinh^6\left(r_1t-q_1\right)
Sinh^2\left(r_2 t -q_2\right)\right] -
6\left(48-\frac{1}{3c_{_\ell}}\right)\nu_\ell\times \\
&\phantom{{}={}} \rm Ln\,\left[Sinh
\left(r_1t-q_1\right)\,Sinh\left(r_2 t-q_2 \right) \right]
\biggr]\rm
+ \frac{3}{\lambda_1}p_3 \left[16\left(72\nu_\ell-1\right)-8\frac{\nu_{_\ell}}{c_{_\ell}}\right]\Delta t, \\
\rm \phi_2 &= \rm\phi_{20}+\frac{1}{2\lambda_2}\biggl[Ln\left[Sinh^2\left(r_1t-q_1\right)
Sinh^6\left(r_2 t-q_2\right)\right] - 6\left(48-\frac{1}{3c_{_\ell}}\right)\nu_\ell \times \\
&\phantom{{}={}} \rm Ln\,\left[Sinh\left(r_1t-q_1\right)\,Sinh\left(r_2 t-q_2 \right) \right]\biggr]\rm + \frac{3}{\lambda_2}p_3
\left[16\left(72\nu_\ell-1\right)-8\frac{\nu_{_\ell}}{c_{_\ell}}\right]\Delta
t,
\end{split}
\end{equation}
where $\eta_0, \phi_{10}$ and $\phi_{20}$ are given in terms of the
$\beta_i$ constants as
\begin{eqnarray}
\eta_0=\frac{\beta_1+\beta_2-6\beta_3}{12},  \label{omega-1}\quad
 \phi_{10}=\frac{3\beta_1+\beta_2-6\beta_3}{2\lambda_1},\label{phi1-1}\quad
 \phi_{20}=\frac{\beta_1+3\beta_2-6\beta_3}{2\lambda_2}.\label{phi2-11}
\end{eqnarray}
It is clear that the standard quintessence model with two scalar
fields cannot be reproduced under this approach, because when we set
$\rm m^{12}=0$, this imply that parameter $\ell$ is equal to zero,
then, the matrix elements $\rm m^{11}=m^{22}$ are zero too, this was
the challenge to resolve.


\section{Quantum Approach}\label{quantum-versions}
Works related to the Wheeler-DeWitt (WDW) equation and the problems
that tackles are extensive, for example in Ref. \cite{Gibbons}, the
question of what a typical wave function for the universe is, is
addressed. In Ref. \cite{Zhi} an excellent summary is presented on
quantum cosmology, where the problem of how the universe emerged
from big bang singularity can no longer be neglected in the GUT
epoch. On the other hand, the best candidates for quantum solutions
become those that have a damping behavior with respect to the scale
factor, since these allow to obtain good classical solutions when
using the WKB approximation for any scenario in the evolution of our
universe \cite{HH,H}.

In this section we present the quantum version of the classical anisotropic
cosmological models studied above along with its solutions. Since we
already have the classical Hamiltonian density, the quantum
counterpart can be obtained making the usual replacement $\rm
\Pi_{q^\mu}=-i\hbar \partial_{q^\mu}$. First we modified the
classical Hamiltonian density (\ref{second-hamibianchi}) in order to
consider the factor ordering problem between the function $\rm
e^{-3\Omega}$ and its moment $\rm \pi_u$, introducing the linear
term as $\rm  e^{-3u}\pi_u^2 \to e^{-3u}\left[\pi_u^2 +Qi\hbar \pi_u
\right]$ where Q is a real number that measures the ambiguity in the
factor ordering.

\subsection{Quantum Anisotropic Quintessence-K-essence Model}
In this section we present the quantum version for the cosmological model studied in Section~\ref{first-model}. We start with the modified
Hamiltonian density,
\begin{equation}
\rm {\cal H}=  \pi_u^2 +Qi\hbar \pi_u-12
 \Pi_{\phi_1}^2 - 12
\Pi_{\phi_2}^2 -6P_1^2-6P_2^2-6P_3^2 -24V_1 e^{6u-\lambda_1\phi_1}
\,, \label{mod-hami-bianchi}
\end{equation}
in order to obtain the WDW equation, we
implement the following change of variables
$\rm(u,\phi_1,\phi_2,u_i)\leftrightarrow (\xi_1,\xi_2,\xi_3)$
\begin{equation}\label{uuno}
\begin{split}
\rm \xi_1 &=\rm6u-\lambda_1 \phi_1, \\
\xi_2&=\rm u, \\
\xi_3&=\phi_2,
\end{split}
\qquad\longleftrightarrow\qquad
\begin{split}
\rm u &=\rm \xi_2, \\
\phi_1&=\frac{-\xi_1+6\xi_2}{\lambda_1}, \\
\phi_2&= \xi_3,\\
\rm u_i&=\rm u_i,
\end{split}
\end{equation}
where $\rm u_i$ are the conjugate coordinate to momenta $\rm P_i$,
and also, obtaining a new set of conjugate momenta (in the same
manner as (\ref{momenta-s})), of the variables $\rm
(\xi_1,\xi_2,\xi_3)$, namely $\rm (\pi_1,\pi_2,\pi_3)$, which read
\begin{equation}
\rm \pi_u=   6 \pi_1 + \pi_2, \qquad  \Pi_{\phi_1} = -\lambda_1
\pi_1, \qquad  \Pi_{\phi_2} = \pi_3, \label{new-momenta}
\end{equation}
which in turn transforms the Hamiltonian density
(\ref{mod-hami-bianchi}) as
\begin{equation}
\rm {\cal H} = \rm  12 \left(3-\lambda_1^2 \right)\pi_1^2+ \pi_2^2
+12 \pi_1 \pi_2-12 \pi_3^2+i\hbar Q(6\pi_1 +
\pi_2)-6P_1^2-6P_2^2-6P_3^2 -  24V_1 e^{\xi_1}. \label{new-hami}
\end{equation}
Introducing the replacement $\rm
\pi_{q^\mu}=-i\hbar\partial_{q^\mu}$, the WDW equation becomes
\begin{align}
\rm {\cal H}\Psi = &-12 \hbar^2\left(3-\lambda_1^2
\right)\frac{\partial^2\Psi}{\partial \xi_1^2}-\hbar^2
\frac{\partial^2\Psi}{\partial \xi_2^2} - 12\hbar^2
\frac{\partial^2\Psi}{\partial \xi_1 \partial \xi_2}+12 \hbar^2
\frac{\partial^2\Psi}{\partial \xi_3^2}+\nonumber\\
&Q\hbar^2
\left(6\frac{\partial \Psi}{\partial \xi_1}+\frac{\partial
\Psi}{\xi_2}\right)+ 6\hbar^2 \left(\frac{\partial^2 \Psi}{\partial
u_1^2}+ \frac{\partial^2 \Psi}{\partial u_2^2}+ \frac{\partial^2
\Psi}{\partial u_3^3}\right) - 24V_1 e^{\xi_1}\Psi=0, \label{q-mod}
\end{align}
due that the scalar potential does not depend on the coordinates
$\rm (\xi_2,\xi_3, u_i)$, we propose the following ansatz for the
wave function $\rm
\Psi(\xi_1,\xi_2,\xi_3,u_i)=e^{-(a_2\xi_2+a_3\xi_3+a_4
u_1+a_5u_2+a_6u_3)/\hbar} G(\xi_1)$ where the $\rm a_i$ are
arbitrary constants. Introducing the mentioned ansatz in
(\ref{q-mod}) we have that
\begin{equation}
\rm -12 \hbar^2\left(3-\lambda_1^2 \right)\frac{1}{G}\frac{d^2G}{d
\xi_1^2} +6  \hbar\left(2 a_2+\hbar Q\right) \frac{1}{G}
\frac{dG}{d\xi_1}-a_2(a_2+\hbar Q)+12 a_3^2 +6a_0^2- 24V_1
e^{\xi_1}=0, \nonumber
\end{equation}
where $ a_0^2=a_4^2+a_5^2+a_6^2$, and also we have divided the whole equation by the ansatz; this in turn
leads us to the following differential equation
\begin{equation}
\rm\frac{d^2G}{d\xi_1^2} -\frac{2a_2+ \hbar
Q}{2\hbar(3-\lambda_1^2)}\frac{dG}{d\xi_1}
+\frac{1}{12\hbar^2(3-\lambda_1^2)} \left[24V_1e^{\xi_1}+ \eta
\right]G=0,\label{qauntum_psi_1}
\end{equation}
here $\eta=a_2(a_2+\hbar \rm Q)-12 a_3^2-6a_0^2$. The last equation can
be casted as $\rm y^{\prime \prime} + a y^\prime + \left(b e^{\kappa
x } +c \right)y=0$ (and whose solutions will depend on the value of
$\lambda_1$) \cite{polyanin}, where
\begin{equation}
\rm y=Exp\left({-\frac{ax}{2}}\right) Z_\nu
\left(\frac{2\sqrt{b}}{\kappa} e^{\frac{\kappa x}{2}} \right),
\end{equation}
here $\rm Z_\nu$ is the Bessel function and
$\nu=\sqrt{a^2-4c}/\kappa$ being the order. The corresponding
relations between the coefficients of (\ref{qauntum_psi_1}) and $\rm
a,b,c$ and $\kappa$ are
\begin{equation}
\rm a =\rm \left\{
\begin{tabular}{lr}
$\rm \frac{2a_2+ \hbar Q}{2\hbar(\lambda_1^2-3)},$ &when $\lambda_1^2 > 3$ \\ \\
$\rm  -\frac{2a_2+ \hbar Q}{2\hbar(3-\lambda_1^2)},$ & when $\lambda_1^2 < 3$
\end{tabular}\right.
\end{equation}
\begin{equation}
\rm b =\rm \left\{
\begin{tabular}{ll}
$\rm -\frac{2V_0}{\hbar^2(\lambda^2-3)},$ & when \,\,$\lambda_1^2 > 3$\\ \\
$\rm \frac{2V_0}{\hbar^2(3-\lambda^2)},$ & when \,\,$\lambda_1^2 <3$
\end{tabular}
\right.
\end{equation}
\begin{equation}
\rm c = \rm \left\{
\begin{tabular}{ll}
$\rm -\frac{\eta}{12 \hbar^2\left(\lambda_1^2-3\right)}$, & when \,\,$\lambda_1^2 > 3$\\ \\
$\rm \frac{\eta}{12\hbar^2\left(3-\lambda_1^2\right)}$,& when
\,\,$\lambda_1^2 < 3$
\end{tabular}
\right.
\end{equation}
\begin{equation}
\kappa=1,
\end{equation}
according to the constant b, the solution to the function G becomes
\begin{align}
\rm G(\xi_1) =\rm Exp\left(-\frac{2a_2+ \hbar
Q}{4\hbar(\lambda_1^2-3)}\xi_1 \right)\,\,\, K_{\nu_1}\left(
\frac{2}{\hbar}\sqrt{\frac{2V_0}{ \lambda^2-3 }}\,\, e^{\frac{\xi_1}{2}} \right), \qquad \lambda_1^2 > 3 \label{k0}\\
\rm G(\xi_1) =\rm Exp\left(\frac{2a_2+ \hbar
Q}{4\hbar(3-\lambda_1^2)\xi_1} \right)\,\,\,
J_{\nu_2}\left(\frac{2}{\hbar}\sqrt{\frac{2V_0}{ 3-\lambda_1^2}}\,\,
 e^{\frac{\xi_1}{2}} \right), \qquad \lambda_1^2 < 3 \label{j0}
\end{align}
and the wave function takes the form
\begin{align}
\rm \Psi_{\nu_1}=&~\rm Exp\left(-\frac{2a_2+ \hbar
Q}{4\hbar(\lambda_1^2-3)}\xi_1 - \frac{a_2 \xi_2 + a_3 \xi_3}{\hbar}
-\frac{a_4u_1+a_4u_2+a_6u_3}{\hbar}\right)\times \nonumber \\
&  K_{\nu_1}\left(\frac{2}{\hbar}\sqrt{\frac{2V_0}{ \lambda_1^2-3 }}\,\, e^{\frac{\xi_1}{2}} \right), \qquad\qquad \lambda_1^2 > 3 \label{k1}\\
\rm \Psi_{\nu_2}=&~\rm  Exp\left(\frac{2a_2+ \hbar
Q}{4\hbar(3-\lambda_1^2)}\xi_1 - \frac{a_2 \xi_2 + a_3 \xi_3}{\hbar}
-\frac{a_4u_1+a_4u_2+a_6u_3}{\hbar}\right)\times \nonumber \\
& J_{\nu_2}\left(\frac{2}{\hbar}\sqrt{\frac{2V_0}{ 3-\lambda^2}}\,\,
 e^{\frac{\xi_1}{2}} \right), \qquad\qquad \lambda_1^2 < 3. \label{j1}
\end{align}
where $\rm \nu_1=\sqrt{\left(-\frac{2a_2+ \hbar
Q}{4\hbar(\lambda_1^2-3)} \right)^2 +
\frac{4\eta}{12\hbar^2(\lambda_1^2-3)}}$ and
 $\rm \nu_2=\sqrt{\left(\frac{2a_2+ \hbar
Q}{4\hbar(3-\lambda_1^2)} \right)^2 -
\frac{4\eta}{12\hbar^2(3-\lambda_1^2)}}$ are the corresponding
order of $\Psi$. Applying the inverse transformation on
the variables $\rm \xi_i$, we can write the wave function in terms
of the original variables $\rm (A=e^\Omega,\phi_i, m_i=e^{u_i})$, which
read
\begin{align}
\rm \Psi_{\nu_1} =&~\rm
m_1^{-\frac{a_4}{\hbar}}m_2^{-\frac{a_5}{\hbar}}m_3^{-\frac{a_6}{\hbar}}
A^{-\alpha_1 }\,Exp\left(\frac{2a_2+ \hbar
Q}{4\hbar(\lambda_1^2-3)}\lambda_1 \phi_1 - \frac{ a_3}{\hbar}\phi_2
\right)\times\nonumber\\
& K_{\nu_1}\left(
\frac{2}{\hbar}\sqrt{\frac{2V_0}{ \lambda_1^2-3 }}\,\, A^{3}e^{\frac{\lambda_1}{2}\phi_1} \right), \qquad\qquad \lambda_1^2 > 3 \label{k2}\\
\rm \Psi_{\nu_2} =&~\rm
m_1^{-\frac{a_4}{\hbar}}m_2^{-\frac{a_5}{\hbar}}m_3^{-\frac{a_6}{\hbar}}A^{-\alpha_2
}\, Exp\left(-\frac{2a_2+ \hbar Q}{4\hbar(\lambda_1^2-3)}\lambda_1
\phi_1 - \frac{ a_3}{\hbar}\phi_2 \right)\times\nonumber \\
&J_{\nu_2}\left(\frac{2}{\hbar}\sqrt{\frac{2V_0}{ 3-\lambda^2}}\,\,
 A^{3}e^{\frac{\lambda_1}{2}\phi_1} \right), \qquad\qquad \lambda_1^2 < 3. \label{j2}
\end{align}
with $\alpha_1=\frac{1}{\hbar}\left( a_2+\frac{3}{2}\frac{2a_2+\hbar
Q}{\lambda_1^2-3}\right)$ and $\alpha_2=\frac{1}{\hbar}\left(
a_2-\frac{3}{2}\frac{2a_2+\hbar Q}{3-\lambda_1^2}\right)$.
\begin{figure}[ht!]
\begin{center}
\includegraphics[scale=0.5]{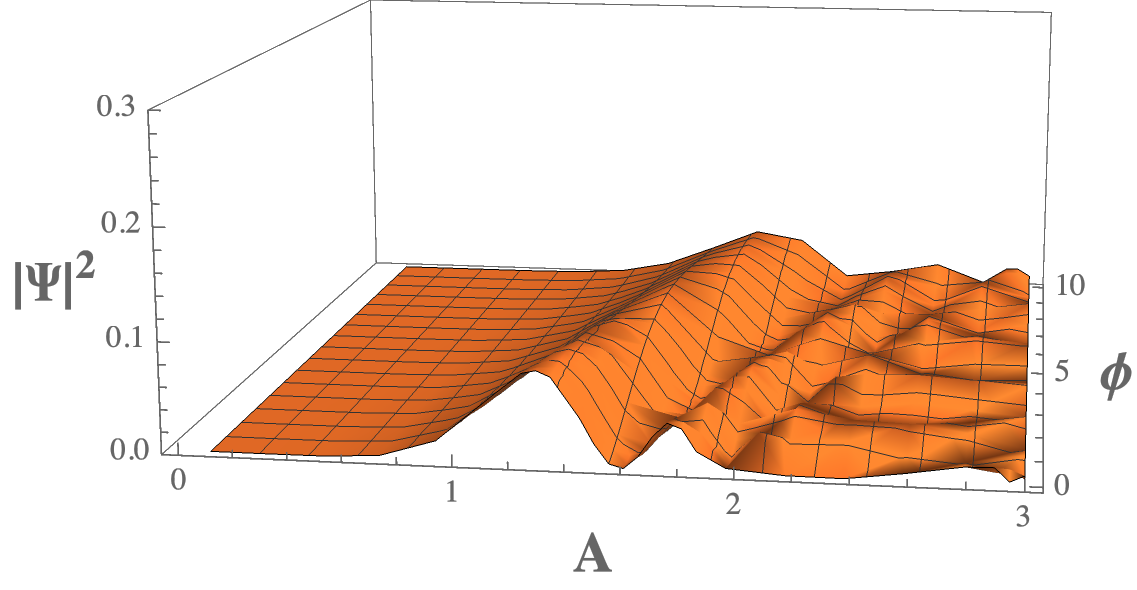}
\caption{Behavior of the probability density for $\lambda_1<\sqrt{3}$, for $\rm Q=1$, $\lambda_1$, $a_2=0.6$, $a_3=1$ and $a_4=a_5=a_6=0.3$.}
\label{density_Q1}
\end{center}
\end{figure}
In Fig(\ref{density_Q1}) we can see the behavior of the probability density of the wave function for the solution $\lambda_1<\sqrt3$. It is
observed that the evolution of the wave function with respect of the scale factor is damped, which is a good characteristic and this kind of
behavior also have been reported in \cite{sor,Socorro:2020nsm,Socorro:2018amv}. In comparison with isotropic model \cite{Socorro:2020nsm},
we can see that the anisotropies shrink the probability density of the wave function.
\begin{figure}[ht!]
\begin{center}
\includegraphics[scale=0.33]{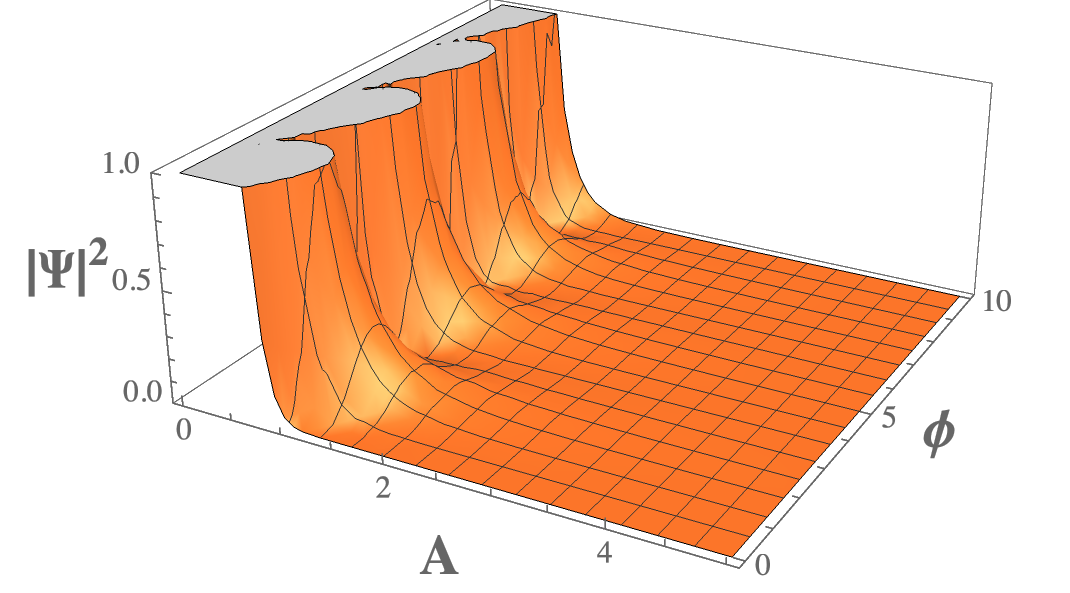}
\includegraphics[scale=0.33]{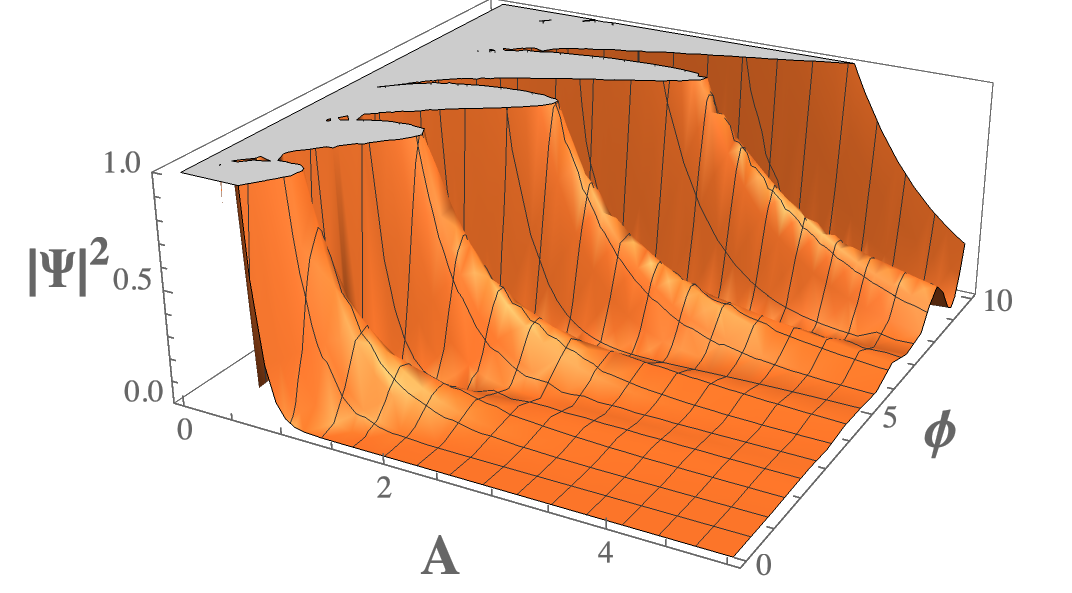}
\caption{Behavior of the probability density for $\lambda_1>\sqrt3$. For both figures $\lambda_1=6$, $a_2=2$, $a_3=1$, $a_4=a_5=a_6=0.3$
 whereas for the left figure $\rm Q=-2$ and for the right figure $\rm Q=-10$.}\label{density_Q2_Q10}
\end{center}
\end{figure}

In Fig.(\ref{density_Q2_Q10}) we can observe the evolution of the wave function for the solution $\lambda_1>\sqrt3$. In this particular
case the values of $\rm Q$ act as a retarder (for negative values) for the wave function and compresses the length over the axis were the
scalar field evolves (this should also delay the inflationary epoch), but still having the damped behavior. Contrasting this results with
those of the isotropic treatment \cite{Socorro:2020nsm}, we can see that anisotropies shrink the the probability density along the evolution
of the scalar field.

Finally, for the particular case of $\lambda_1=\sqrt3$ the quantum solution for
the function $\rm G(\xi_1)$ becomes
$$\rm G(\xi_1)=G_0 Exp\left[\frac{\eta}{6\hbar(2a_2+\hbar Q)}\xi_1 \right]\, Exp\left(\frac{4V_1}{\hbar(2a_2+\hbar Q)} e^{\xi_1}
\right),$$ and the wave function is
\begin{align}\label{sol_139}
\rm \Psi(A,\phi_i,m_i)=&~\rm \Psi_0
m_1^{-\frac{a_4}{\hbar}}m_2^{-\frac{a_5}{\hbar}}m_3^{-\frac{a_6}{\hbar}}
A^{r}\,Exp\left(-\frac{a_3}{\hbar}\phi_2-\frac{\lambda_1}{6\hbar(2a_2+\hbar
Q)} \phi_1 \right)\times\nonumber\\
&\rm Exp\left[ \frac{4V_1}{\hbar(2a_2+\hbar Q)}A^6
e^{-\lambda_1 \phi_1})\right].
\end{align}
where the constant $\rm
r=-\frac{a_2}{\hbar}+\frac{\eta}{\hbar(2a_2+\hbar Q)}$.

\subsection{Quantum Anisotropic Quintom Case}\label{quantum_quintom}
For the second cosmological model, the quintom like case, the
quantum version of this model is obtained applying, again, the
recipe $\rm \Pi_{q^\mu}=-i\hbar \partial_{q^\mu}$ to the Hamiltonian
density (\ref{hami-bianchi-q}), hence
\begin{equation}\label{q-hami-bianchi-qu}
\begin{split}
\rm \biggl[ \frac{\hbar^2}{\mu_{_\ell}}\frac{\partial^2}{\partial
\xi_1^2} +\frac{\hbar^2}{\mu_{_\ell}}\frac{\partial^2}{\partial
\xi_2^2} -\hbar^2\left(48-\frac{1}{3c_{_\ell}}
\right)\left(\frac{\partial^2}{\partial \xi_3 \partial \xi_1} +
\frac{\partial^2}{\partial \xi_3 \partial \xi_2}
 \right)
-\hbar^2\left(16-\frac{1}{18c_{_\ell}}\right)\frac{\partial^2}{\partial
\xi_3^2}\\
+6\hbar^2\left(\frac{\partial^2 \Psi}{\partial
u_1^2}+\frac{\partial^2 \Psi}{\partial u_2^2}+\frac{\partial^2
\Psi}{\partial u_3^2}\right)-24V_1e^{-\xi_1}-24V_2 e^{-\xi_2}\biggr]
\Psi=0,
\end{split}
\end{equation}
because the scalar potential does not depend on the coordinate $\rm
\xi_3$, we propose the following ansatz for the wave function $\rm
\Psi(\xi_1,\xi_2,\xi_3)=e^{(a_3\xi_3+a_4u_1+a_4u_2+a_6u_3)/\hbar}\,{\cal
A}(\xi_1) {\cal B}(\xi_2)$ where $\rm a_i$ (with $\rm i=3,4, 5, 6$) are an
arbitrary constants. Substituting and dividing by the ansatz in
(\ref{q-hami-bianchi-qu}), we obtain
\begin{align}
\rm \frac{\hbar^2}{\mu_{_\ell}{\cal A}} \frac{d^2 {\cal
A}}{d\xi_1^2}+ \frac{\hbar^2}{\mu_{_\ell}{\cal B}} \frac{d^2 {\cal
B}}{d\xi_2^2} & -a_3 \hbar \left(48-\frac{1}{3c_{_\ell}}
\right)\left(\frac{1}{{\cal A}}\frac{d{\cal A}}{d \xi_1} +
\frac{1}{{\cal B}} \frac{d{\cal B}}{d \xi_2}
\right)\nonumber\\
&-a_3^2\left(16-\frac{1}{18c_{_\ell}}\right)
+6a_0^2-24V_1e^{-\xi_1}-24V_2 e^{-\xi_2}=0,
\end{align}
with $\rm a_0^2=a_4^2+a_5^2+a_6^2$, where we can separate the
equations as
\begin{align}
\rm \frac{d^2 {\cal A}}{d\xi_1^2}- & \frac{a_3 \mu_{_\ell}}{\hbar}
\left(48-\frac{1}{3c_{_\ell}} \right) \frac{d{\cal A}}{d \xi_1}\nonumber\\
&-\frac{\mu_{_\ell}}{\hbar^2} \left(\frac{a_3^2}{2}
\left(16-\frac{1}{18c_{_\ell}}\right)-3a_0^2-\alpha^2+24V_1e^{-\xi_1}
\right) {\cal A}=0, \label{a}\\
\rm \frac{d^2 {\cal B}}{d\xi_2^2}-&\frac{a_3 \mu_{_\ell} }{\hbar}
\left(48-\frac{1}{3c_{_\ell}} \right)  \frac{d{\cal B}}{d \xi_2}\nonumber\\
&-\frac{\mu_{_\ell}}{\hbar^2}\left(\frac{a_3^2}{2}
\left(16-\frac{1}{18c_{_\ell}}\right)-3a_0^2+\alpha^2+24V_2e^{-\xi_2}\right)
{\cal B}=0, \label{b}
\end{align}
with $\alpha^2$ being the separation constant. The corresponding
solutions of Eqs. (\ref{a}) and (\ref{b}) have the following form
\cite{polyanin}
\begin{equation}
\rm Y(x)=Exp\left({-\frac{ax}{2}}\right)
Z_\nu\left(\frac{2\sqrt{b}}{\lambda} e^{\frac{\lambda x}{2}}\right),
\end{equation}
here $\rm Z_\nu$ are the generic Bessel function with order $\rm
\nu=\sqrt{a^2-4c}/\lambda$. If $\sqrt{b}$ is real, $\rm Z_\nu$ are
the ordinary Bessel function, otherwise the solution will be given
by the modified Bessel function. Making the following
identifications
\begin{eqnarray}
\rm \lambda&=&-1, \\
\rm a&=&-\frac{a_3 \mu_{_\ell}}{\hbar}\left(48-\frac{1}{3c_{_\ell}} \right), \\
\rm b_{1,2}&=&-\frac{\mu_{_\ell}}{\hbar^2}24V_{1,2}, \\
\rm c_\mp&=&-\frac{\mu_{_\ell}}{\hbar^2}\left(a_3^2\left(8-\frac{1}{36c_{_\ell}}\right)-3a_0^2 \mp \alpha^2\right),\\
\rm \nu_\mp&=&\sqrt{\frac{\rm a^2}{\mu_\ell}+4\rm c_\mp},
\end{eqnarray}
we can check that the value for $\sqrt{b}$ is imaginary, which as
already mentioned, gives a solution in terms of the modified Bessel
function $\rm Z_\nu=K_\nu$ whose order lies in the reals. Thus, the
wave function is
\begin{align}\label{qsol_quintom}
\rm \Psi_{\nu_\pm}=&~\rm Exp\left[\left(\frac{\mu_{_\ell}}{2\hbar}\left(48-\frac{1}{3c_{_\ell}}\right)(\xi_1+\xi_2)+\frac{\xi_3}{\hbar}+
\frac{a_4u_1+a_5u_2+a_6u_3}{\hbar}\right)a_3\right]\times\nonumber\\
&\rm K_{\nu_-}\left(\frac{4}{\hbar}\sqrt{6V_1\mu_{_\ell}}
e^{-\frac{\xi_1}{2}} \right)
K_{\nu_+}\left(\frac{4}{\hbar}\sqrt{6V_2\mu_{_\ell}}
e^{-\frac{\xi_2}{2}} \right).
\end{align}

\subsection{Quantum Anisotropic Quintessence Case}
Lastly, we are going to consider the quantum version of the
anisotropic quintessence like case. As in the previous two subsections, what we want
is to obtain an equation of the form $\rm {\cal H}\Psi(\xi_i)=0$, to
achieve this we introduce the standard prescription
$\Pi_q^\mu=-i\hbar\partial_{q^\mu}$ in (\ref{hami-bianchi-qi}),
obtaining
\begin{eqnarray}
&&\rm \left[ -\frac{\hbar^2}{\nu_{_\ell}}\frac{\partial^2}{\partial
\xi_1^2} -\frac{\hbar^2}{\nu_{_\ell}}\frac{\partial^2}{\partial
\xi_2^2} -\hbar^2\left(24+\frac{1}{3\nu_{_\ell}}
\right)\left(\frac{\partial^2}{\partial \xi_3 \partial \xi_1} +
\frac{\partial^2}{\partial \xi_3 \partial \xi_2}
 \right)
-\hbar^2\left(12+\frac{1}{18\nu_{_\ell}}\right)\frac{\partial^2}{\partial
\xi_3^2} \right.\nonumber\\
&& \left. \rm +6\hbar^2 \left(\frac{\partial^2 \Psi}{\partial
u_1^2}+\frac{\partial^2 \Psi}{\partial u_2^2}+\frac{\partial^2
\Psi}{\partial u_3^2}\right)-24V_1 e^{-\xi_1}-24V_2
e^{-\xi_2}\right] \Psi=0, \label{q-hami-bianchi-qi}
\end{eqnarray}
we can see that the scalar potential does not depend on the
coordinates $\rm \xi_3,u_i$, consequently we propose the following
ansatz for the wave function $\rm
\Psi(\xi_1,\xi_2,\xi_3)=e^{(b_3\xi_3+b_4u_1+b_5u_2+b_6u_3)/\hbar}\,
{\cal A}(\xi_1) {\cal B}(\xi_2)$ where $\rm b_i$ (with $\rm i=3,4, 5, 6$) are an arbitrary
constant. Applying and dividing by the ansatz in
(\ref{q-hami-bianchi-qi}) we get
\begin{align}
& \rm -\frac{\hbar^2}{\nu_{_\ell}{\cal A}} \frac{d^2 {\cal
A}}{d\xi_1^2}- \frac{\hbar^2}{\nu_{_\ell}{\cal B}} \frac{d^2 {\cal
B}}{d\xi_2^2}- b_3 \hbar \left(48-\frac{1}{3c_{_\ell}}
\right)\left(\frac{1}{{\cal A}}\frac{d{\cal A}}{d \xi_1} +
\frac{1}{{\cal B}} \frac{d{\cal B}}{d \xi_2} \right) -b_3^2
\left(16-\frac{1}{18c_{_\ell}}\right)\nonumber\\
&+6b_0^2 -24V_1e^{-\xi_1}-24V_2
e^{-\xi_2}=0,
\end{align}
with $\rm b_0^2=b_4^2+b_5^2+b_6^2$, separating the equations we have
that
\begin{align}\label{bb}
\rm \frac{d^2 {\cal A}}{d\xi_1^2}&+ \frac{b_3 \nu_{_\ell}}{\hbar}
\left(48-\frac{1}{3c_{_\ell}} \right)\frac{d{\cal A}}{d \xi_1}\\\nonumber
&+\frac{\nu_{_\ell}}{\hbar^2} \left(b_3^2
\left(8-\frac{1}{36c_{_\ell}}\right)-3b_0^2-\alpha^2+24V_1e^{-\xi_1}
\right) {\cal A}=0,\\
\rm \frac{d^2 {\cal B}}{d\xi_2^2}&+\frac{b_3 \nu_{_\ell} }{\hbar}
\left(48-\frac{1}{3c_{_\ell}} \right)\frac{d{\cal B}}{d \xi_2}\\\nonumber
&+\frac{\mu_{_\ell}}{\hbar^2}\left(b_3^2
\left(8-\frac{1}{36c_{_\ell}}\right)-3b_0^2+\alpha^2+24V_2e^{-\xi_2}\right)
{\cal B}=0,
\end{align}
where $\alpha^2$ is the separation constant. These last two
equations are similar to those of the quantum quintom like case (\ref{a}) and
(\ref{b}). Proceeding in a similar fashion as the previous
subsection (\ref{quantum_quintom}), we make the following
identifications
\begin{eqnarray}
\rm \lambda&=&-1, \\
\rm a&=&\frac{b_3 \nu_{_\ell}}{\hbar}\left(48-\frac{1}{3c_{_\ell}} \right), \\
\rm b_{1,2}&=&\frac{\nu_{_\ell}}{\hbar^2}24V_{1,2}, \\
\rm c_\mp&=&\frac{\nu_{_\ell}}{\hbar^2}\left(b_3^2\left(8-\frac{1}{36c_{_\ell}}\right)- 3a_0^2 \mp \alpha^2\right), \\
\end{eqnarray}
and conclude that the solutions are given by the ordinary Bessel
function $J_\nu$ with order $\rm\nu_\mp=\sqrt{(\rm
a^2/\nu_\ell)+4\rm c_\mp}$. Thus, the wave function becomes
\begin{align}
\rm \Psi_{\nu_\pm}=&~\rm Exp\left[\left(\frac{\nu_{_\ell}}{2\hbar}\left(48-\frac{1}{3c_{_\ell}}\right)(-\xi_1-\xi_2)
+\frac{\xi_3}{\hbar}\right)b_3+\frac{b_4u_1+b_5u_2+b_6u_3}{\hbar}\right]\times\nonumber\\
&\rm J_{\nu_-}\left(\frac{4}{\hbar}\sqrt{6V_1\nu_{_\ell}}e^{-\frac{\xi_1}{2}} \right)
J_{\nu_+}\left(\frac{4}{\hbar}\sqrt{6V_2\nu_{_\ell}}e^{-\frac{\xi_2}{2}} \right).\label{qsol_quintessence}
\end{align}
\section{Final Remarks}\label{conclusions}

In this work we have studied the anisotropic Bianchi type model in the chiral cosmology setup in a twofold way.
In the first cosmological model we consider two scalar fields but a single term of the potential. In the second one, additionally to the two
scalar fields, we also consider both terms in the potential. For both models we did a classical and quantum treatment, obtaining exact analytical
solutions for both scenarios.

In the first model, which can be thought as a quintessence plus k-essence model, our findings show that the volume of the
universe grows in an accelerated manner for each of the three exact solutions that were found. This feature can be seen from
Fig.(\ref{q-parameters}), where solutions for $\lambda_1<\sqrt3$ and $\lambda_1=\sqrt3$ have a similar behavior whereas
 the solution for $\lambda>\sqrt3$ has a more faster evolution. After a certain amount of time, the three solutions stabilized at the same value of -1.
Also, because of the ratio of the shear to scalar expansion bound:
$\sigma/\theta\leq 0.3$ \cite{Pradhan:2010dm}, we were able to
constrain the value of the anisotropic parameter $\rm \overline A_m$
for two of the solutions, however the anisotropy continue, because
for the cases $\lambda_1>\sqrt{3}$ and $\lambda_1=\sqrt{3}$ the
anisotropic parameter becomes $\rm \overline A_m\leq 3.54$. In the
quantum regime we were also able to find exact solutions. For the
particular
 case of $\lambda_1<\sqrt3$ we found that the wave function has a damped behavior as the scale factor evolves, as can be seen
 in Fig.(\ref{density_Q1}), this distinctive mark have also been reported in  \cite{sor,Socorro:2020nsm,Socorro:2018amv}.
 In contrast with the isotropic treatment \cite{Socorro:2020nsm}, we found that the anisotropies shrink the probability density of the wave function.
  For the solution $\lambda_1>\sqrt3$, it is found that the damped behavior still exists, but the
 parameter $\rm Q$ acts as a retarder (for negative values) for the wave function and the length over the axis were the field
 evolves is compressed as shown in Fig.(\ref{density_Q2_Q10}), signaling that the inflation epoch should also be retarded in time. In this case
 the anisotropies shrink the probability density along the evolution of the scalar field. Finally, equation (\ref{sol_139}) depicts the quantum
 solution for the case $\lambda_1=\sqrt3$.

For the second model under study, we consider both potential terms, in addition to the two scalar fields. In this setup two possible avenues
were distinguished: a quintom one and a quintessence one. Classical exact solutions for the former are given by equations (\ref{sols_quintom}),
while the solutions for the latter are given by equations (\ref{sols_quintessence}). For the quantum counterpart exact solutions were also
obtained. The quantum solutions for both the anisotropic quintom case and the anisotropic quintessence case, are given in terms of exponential
functions (that has the anisotropic information) multiplied by the modified Bessel function $K_{\pm\nu}$ and the ordinary Bessel function
$J_{\pm\nu}$, as depicted in Eq.(\ref{qsol_quintom}) and Eq.(\ref{qsol_quintessence}), respectively.


\acknowledgments{\noindent This work was partially supported by
PROMEP grants UGTO-CA-3. J.S. is partially supported SNI-CONACYT.
This work is part of the collaboration within the Instituto Avanzado
de Cosmolog\'{\i}a and Red PROMEP: Gravitation and Mathematical
Physics under project {\it Quantum aspects of gravity in
cosmological models, phenomenology and geometry of space-time}. Many
calculations where done by Symbolic Program REDUCE 3.8.}


\end{document}